\documentclass[12pt]{article}

\usepackage{graphicx,rotating,epsfig,amsmath}  
\usepackage{multirow}
\newlength{\digitwidth} 

\newcommand{\HRule}{\rule{0.4\linewidth}{0.3mm}}
\newcommand{\mlab}[1]%
    {\mbox{}\marginpar{\raggedright\hspace{0pt}\footnotesize #1}}

\newcommand{\pt}{$p_{\mathrm{T}}$}
\newcommand{\xf}{$x_{\mathrm{F}}$}
\newcommand{\jpsi}{J/$\psi$}
\newcommand{\psip}{$\psi^\prime$}
\newcommand{\ccbar}{$c\overline{c}$}
\newcommand{\degree}{\mbox{$^\circ$}}

\begin{document}

\begingroup
\thispagestyle{empty}
\baselineskip=14pt
\parskip 0pt plus 5pt

\begin{center}
\large EUROPEAN LABORATORY FOR PARTICLE PHYSICS
\end{center}

\bigskip
\begin{flushright}
CERN--PH--EP\,/\,2006-018\\
July 6, 2006
\end{flushright} 

\bigskip\bigskip
\begin{center}
\Large\bf
\jpsi\ and \psip\ production and their normal nuclear absorption in 
proton-nucleus collisions at 400~GeV
\end{center}

\bigskip\bigskip
\begin{center}
\small
\end{center}

\begin{center}
\emph{\large NA50 Collaboration}
\end{center}
\begin{center}
B.~Alessandro$^{10}$,
C.~Alexa$^{3}$,
R.~Arnaldi$^{10}$,
M.~Atayan$^{12}$,
S.~Beol\`e$^{10}$,
V.~Boldea$^{3}$,
P.~Bordalo$^{6,{\rm a}}$,
G.~Borges$^{6}$,
J.~Castor$^{2}$,
B.~Chaurand$^{9}$,
B.~Cheynis$^{11}$,
E.~Chiavassa$^{10}$,
C.~Cical\`o$^{4}$,
M.P.~Comets$^{8}$,
S.~Constantinescu$^{3}$,
P.~Cortese$^{1}$,
A.~De~Falco$^{4}$,
N.~De~Marco$^{10}$,
G.~Dellacasa$^{1}$,
A.~Devaux$^{2}$,
S.~Dita$^{3}$,
J.~Fargeix$^{2}$,
P.~Force$^{2}$,
M.~Gallio$^{10}$,
C.~Gerschel$^{8}$,
P.~Giubellino$^{10}$,
M.B.~Golubeva$^{7}$,
A.~Grigoryan$^{12}$,
J.Y.~Grossiord$^{11}$,
F.F.~Guber$^{7}$,
A.~Guichard$^{11}$,
H.~Gulkanyan$^{12}$,
M.~Idzik$^{10,{\rm b}}$,
D.~Jouan$^{8}$,
T.L.~Karavicheva$^{7}$,
L.~Kluberg$^{9}$,
A.B.~Kurepin$^{7}$,
Y.~Le~Bornec$^{8}$,
C.~Louren\c co$^{5}$,
M.~Mac~Cormick$^{8}$,
A.~Marzari-Chiesa$^{10}$,
M.~Masera$^{10}$,
A.~Masoni$^{4}$,
M.~Monteno$^{10}$,
A.~Musso$^{10}$,
P.~Petiau$^{9}$,
A.~Piccotti$^{10}$,
J.R.~Pizzi$^{11,{\rm d}}$,
F.~Prino$^{10}$,
G.~Puddu$^{4}$,
C.~Quintans$^{6}$,
L.~Ramello$^{1}$,
S.~Ramos$^{6,{\rm a}}$,
L.~Riccati$^{10}$,
H.~Santos$^{6}$,
P.~Saturnini$^{2}$,
E.~Scomparin$^{10}$,
S.~Serci$^{4}$,
R.~Shahoyan$^{6,{\rm c}}$,
M.~Sitta$^{1}$,
P.~Sonderegger$^{5,{\rm a}}$,
X.~Tarrago$^{8}$,
N.S.~Topilskaya$^{7}$,
G.L.~Usai$^{4}$,
E.~Vercellin$^{10}$,
L.~Villatte$^{8}$,
N.~Willis$^{8}$
\end{center}

\begin{center}
\bigskip
\textbf{Abstract}
\end{center}

\bigskip
\begingroup
We report a new measurement of \jpsi, \psip\ and Drell-Yan
cross-sections, in the kinematical domain $-0.425< y_{\rm cm} <0.575$ 
and $-0.5< \cos \theta_{\rm CS} <0.5$, performed at the CERN-SPS using 
400~GeV/c incident protons on Be, Al, Cu, Ag, W and Pb targets.
The dependence of the charmonia production cross-sections on the size 
of the target nucleus allows to quantify the so-called normal nuclear 
absorption. In the framework of the Glauber model, this new measurement 
is combined with results previously obtained with the same apparatus, 
under different experimental conditions, and leads to a precise 
determination of the \jpsi\ and \psip\ absorption cross-sections in the 
surrounding nuclear matter.
\endgroup

\bigskip
\begin{center}
\emph{Accepted for publication in Euro. Phys. J. C.}
\end{center}

\newpage
\bigskip

\setlength{\parindent}{0mm}
\setlength{\parskip}{0mm}
\small
\HRule\\
\begin{flushleft}

$^{~1}$ Universit\`a del Piemonte Orientale, Alessandria and INFN-Torino, Italy\\
$^{~2}$ LPC, Univ. Blaise Pascal and CNRS-IN2P3, Aubi\`ere, France\\
$^{~3}$ IFA, Bucharest, Romania\\
$^{~4}$ Universit\`a di Cagliari/INFN, Cagliari, Italy\\
$^{~5}$ CERN, Geneva, Switzerland\\
$^{~6}$ LIP, Lisbon, Portugal\\
$^{~7}$ INR, Moscow, Russia\\
$^{~8}$ IPN, Univ. de Paris-Sud and CNRS-IN2P3, Orsay, France\\
$^{~9}$ LLR, Ecole Polytechnique and CNRS-IN2P3, Palaiseau, France\\
$^{10}$ Universit\`a di Torino/INFN, Torino, Italy\\
$^{11}$ IPN, Univ. Claude Bernard Lyon-I and CNRS-IN2P3, Villeurbanne, France\\
$^{12}$ YerPhI, Yerevan, Armenia \\ 

a) also at IST, Universidade T\'ecnica de Lisboa, Lisbon, Portugal\\
b) also at Faculty of Physics and Nuclear Techniques, AGH University of Science and Technology, Cracow, Poland\\
c) on leave of absence of YerPhI, Yerevan, Armenia\\
d) deceased \\
\end{flushleft}
\endgroup

\newpage
\pagenumbering{arabic}
\setcounter{page}{1}

\section{Introduction}

Charmonium production in proton-nucleus collisions has already been
extensively studied.  Several fixed target experiments
(NA3~\cite{Bad83}, E772~\cite{Ald91}, NA38~\cite{MAbreu},
E866~\cite{Lei00}, E672/E706~\cite{Gri00} and 
NA50~\cite{pcpaper,rspaper}) have studied \jpsi\ and
\psip\ production, including the nuclear dependence of their
production cross-sections, in the energy range $\sqrt{s}\sim
20$--40~GeV and as a function of several kinematical variables:
Feynman $x$ ($x_{\mathrm{F}}$), rapidity ($y$) and transverse momentum
(\pt).  These measurements have been analysed in view of understanding
the physics mechanisms underlying charmonia production~\cite{Vogt,Bram}.

The study of \jpsi\ production in proton-nucleus collisions is also
crucial for a correct interpretation of the \jpsi\ suppression
patterns experimentally observed in heavy-ion collisions, at the
CERN/SPS~\cite{anomalous}, predicted 20 years ago to signal Quark
Gluon Plasma formation~\cite{Satz}.  It is, indeed, very important to
establish a robust baseline on the basis of proton-nucleus data, the
so-called ``normal nuclear absorption'', with respect to which we can
clearly identify new phenomena, specific of high-energy heavy-ion
collisions.

This work reports new measurements on \jpsi\ and \psip\ production
from proton-nucleus data collected by the NA50 experiment with 400~GeV
protons incident on various targets.  Together with previous
proton-nucleus results from the same experiment at 450~GeV, this new
determination leads to a good characterization of the charmonia
survival probability in ``cold nuclear matter'' and precisely reveals
the differences between the two charmonium states, \jpsi\ and \psip.

\section{Experimental apparatus and data selection}

The data samples analysed in this paper have been collected with the
muon spectrometer used by the NA50 experiment since 1994, as described
in~\cite{anomalous}. The target region of the experiment, however, has
been adapted to the use of six different target materials,
sequentially exposed to the beam in relatively short time intervals,
as shortly summarized hereafter.

The six different targets chosen for these measurements were made of
Be, Al, Cu, Ag, W and Pb, so as to cover a wide range of nuclear
sizes.  They were disks of 12~mm diameter, much larger than the proton
beam spot of 500~$\mu$m r.m.s., ensuring that no incoming proton
misses the target transverse area. Their thicknesses ($L_{\rm tgt}$)
were chosen in order to yield event samples of similar statistics for
each target species, in data collection periods (``runs'') of around
1--2 hours each, while avoiding too thick targets (more than 13~cm or
40\,\% of an interaction length), which could degrade the
measurements.  Table~\ref{tab:targets} summarizes some target
properties, the numbers in parentheses giving the relative errors
associated with each quantity.  The target densities, $\rho$, were
measured, while the interaction lengths, $\lambda_{\rm int}$, were
computed using the cross-sections published in~\cite{Carroll79}.  The
target effective length is defined as \mbox{$L_{\rm eff}=\lambda_{\rm
int} \left[ 1-\exp{(-\rho L_{\rm tgt}/ \lambda_{\rm int})} \right]$}.

\begin{table}[ht]
\begin{center}
\catcode`?=\active \def?{\kern\digitwidth}
\begin{tabular}{lcccccc}
\hline
\rule{0pt}{0.45cm}
   & $A$ & $L_{\rm tgt}$~(cm) & $\rho$~(g/cm$^3$)
   & $\lambda_{\rm int}$~(g/cm$^2$) & $L_{\rm eff}$~(g/cm$^2$)
   & $L_{\rm eff}$/$\lambda_{\rm int}$ \\
\hline
 Be  & ??9.012 & 13.0 & ?1.845 (0.5\,\%) & ?80.0 (1.3\,\%) & 20.7 (0.5\,\%) & 0.26 \\
 Al  & ?26.982 & 12.0 & ?2.695 (0.5\,\%) & 108.9 (0.5\,\%) & 28.0 (0.4\,\%) & 0.26 \\
 Cu  & ?63.546 & ?7.5 & ?8.92? (0.5\,\%) & 138.6 (0.4\,\%) & 53.1 (0.4\,\%) & 0.38 \\
 Ag  & 107.868 & ?7.5 & 10.489 (0.5\,\%) & 160.9 (0.7\,\%) & 62.2 (0.4\,\%) & 0.39 \\
 W   & 183.84? & ?4.5 & 19.21? (0.5\,\%) & 186.9 (1.0\,\%) & 69.2 (0.4\,\%) & 0.37 \\
 Pb  & 207.2?? & ?6.7 & 11.26? (0.8\,\%) & 193.3 (1.1\,\%) & 62.5 (0.7\,\%) & 0.32 \\
\hline
\end{tabular}
\end{center}
\vglue-2mm
\caption{Properties of the targets used to collect the data samples
  analysed in the present work.}
\label{tab:targets}
\end{table}

A remotely operated rotating target holder placed each different
target in the beam sequentially, with a target change every hour or
so.  This feature allowed us to balance time-dependent effects among
different exposures, reducing to a negligible level the systematic
uncertainties related to beam characteristics.

The hadron absorber separating the target region from the muon
spectrometer was made of 60~cm of Al$_2$O$_3$, followed by 400~cm of
carbon and 80~cm of iron.  The overall beam time allocated to the
measurements reported in this paper was only 4 days, forcing us to run
with a high intensity proton beam (3--$4\times 10^9$ protons per 3.2~s
burst). Nevertheless, the data acquisition worked with average
lifetimes between 99\,\% and 96\,\%, depending on the 
specific trigger rate of each target exposure. The time stability 
of the targeting efficiency was continuously monitored by three 
scintillator telescopes pointing to
the target, with axes orientation at 90\degree\ relative to the beam
line, one vertical and two horizontal.  
The intensity of the incident proton beam was redundantly
measured, for continuous cross-checking purposes, by three independent
ionization chambers placed upstream of the target, with a sensitive 
transverse size which ensured that they intercepted all the incoming 
beam. The ionization current is integrated during every burst and the 
total collected charge is proportional to the number of crossing protons. 
These detectors have been precisely calibrated in a dedicated low intensity 
data collection using a coincidence of two scintillators equipped with very 
fast electronics. Their linearity was checked up to the highest beam intensity 
used in the experiment and a 3\% systematic uncertainty on the number of 
incident protons has been estimated \cite{GborgesPhD}. 

Two measurements of the dimuon trigger efficiency, made in dedicated
runs at the beginning and middle of the data taking period, gave
($86.3\pm0.3$)\,\% and ($88.0\pm0.3$)\,\%, respectively. From the
study of several different observables as a function of time, 
a 2.3\,\% systematic error has been deduced.

The reconstruction efficiency depends on the occupancy of 
the spectrometer Multi Wire Proportional Chambers which is a function 
of the luminosity of each exposure (target length and beam intensity). 
It was determined, with its corresponding error, from the individual 
efficiencies of each of the two sets of 12 wire planes detecting 
uncurved tracks, upstream and downstream from the magnet, and cross-checked 
with a Monte-Carlo study in which generated dimuons embedded into real events 
are processed through the same reconstruction algorithm as applied to 
real data \cite{RubenPhD}. The final absolute experimental values vary 
between 77\% and 84\%, with a 0.2\% systematic uncertainty, depending on 
the individual short time exposure.
%

The dimuon events are selected according to the standard quality
criterium used in NA50, mostly driven by geometrical arguments.  In
particular, each single muon must cross the air core toroidal magnet 
(ACM) of the spectrometer in one of the six air-sectors, and it must 
point to the target. The latter condition is assured by a cut on the 
$p \times D_{\rm target}$ variable, where $p$ is the momentum of the 
muon and $D_{\rm target}$ is the transverse distance, in 
the plane containing the centre of the target, between the beam line 
and the intercept of the extrapolated muon track. Phase space regions 
where the acceptance is less than 10\,\% of its maximum value are
removed by applying the following kinematical cuts: $-0.425< y_{\rm
cm} <0.575$ and $-0.5< \cos \theta_{\rm CS} <0.5$, where
$y_{\rm cm}$ is the dimuon center of mass rapidity and
$\theta_{\rm CS}$ is the polar angle in the Collins-Soper frame.

A small fraction of dimuons from collisions upstream or downstream of
the target survives the selection criteria.  They are estimated from
data samples collected with no target in place, after proper
normalisation to the luminosity integrated in the standard runs.  
The contribution of such spurious dimuons in the mass region
\mbox{$2.7<M<3.5$~GeV/$c^2$} is less than $\sim$\,3\,\% for the lightest
target and less than $\sim$\,1\,\% for the heaviest one. In spite 
of their small number, they must be carefully taken into account 
because the reconstruction program assumes that these dimuons are 
produced in the target and, therefore, their reconstructed invariant 
mass is shifted to regions where they may contribute more significantly 
to the total number of events (in particular at around 2.8~GeV/$c^2$, 
as seen in Fig.~\ref{ppbfit} below).

\begin{table}[ht]
\begin{center}
\catcode`?=\active \def?{\kern\digitwidth}
\begin{tabular}{lccc}
\hline
\rule{0pt}{0.45cm}
 Data set &  $N^{\rm tot}_{\rm protons}$ 
             &  $\langle \epsilon_{\rm trigger}\cdot\epsilon_{\rm rec}\rangle$ 
             &  $N^{+-}_{\mu\mu}$ \\
             &  ($\times 10^{12}$) & & (2.7--3.5) \\
\hline
 p-Be      & 6.355 & 0.736 & 38\,627 \\
 p-Al      & 6.522 & 0.719 & 48\,614 \\
 p-Cu      & 3.496 & 0.686 & 44\,984 \\
 p-Ag      & 2.979 & 0.680 & 41\,560 \\
 p-W       & 3.380 & 0.681 & 49\,476 \\
 p-Pb      & 5.258 & 0.691 & 69\,388 \\
\hline
\end{tabular}
\end{center}
\vglue-2mm
\caption{Some features of the collected data samples.}
\label{datasets}
\end{table}

Table~\ref{datasets} presents, for each data sample, the total number
of incident protons, the average efficiency (affected by a 2.3\,\% 
systematic uncertainty), and the number of reconstructed 
opposite-sign muon pairs in the mass region $2.7<M<3.5$~GeV/$c^2$ 
after the selection cuts.

\section{Data analysis}

The opposite-sign muon pair invariant mass distributions are analysed
as a superposition of several contributions. For masses lower than
2~GeV/$c^2$, the mass spectrum is dominated by uncorrelated decays
of pions and kaons. This background can be estimated from the number
of measured like-sign pairs, according to the standard formula:
$$N^{+-}_{\rm bg} = 2 R_{\rm bg} \left [ \left ( \sqrt{N^{++} N^{--}}
\right )_{\rm ACM +} +\left ( \sqrt{N^{++} N^{--}} \right )_{\rm ACM
-} \right ] \quad, $$ where ACM$+$ and ACM$-$ denote the magnetic
field polarities used to collect the events.  This relation is valid
if the probability to detect a muon in the apparatus (acceptance) is
independent of its charge.  An additional offline cut is therefore
imposed which rejects events where one (or both) of the muon tracks
would not be accepted if it had the opposite charge (or if the
magnetic field had the opposite polarity).  $R_{\rm bg}$ accounts for
charge correlations.  It tends to unity in events of very high charged
track multiplicities, where such correlations can be neglected.  In
the case of our data samples, $R_{\rm bg}$ varies between 1.03 and
1.05, depending on the target, as estimated by fitting the mass region
1.5--2.3~GeV/$c^2$, where the combinatorial background is dominant.
These values are very close to unity because the beam intensities we
have used are so high that it frequently happens that the two
muons of a given triggered event come from a different p-A collision
and are, therefore, uncorrelated.

\begin{figure}[ht]
\centering
\resizebox{0.6\textwidth}{!}{%
\includegraphics*{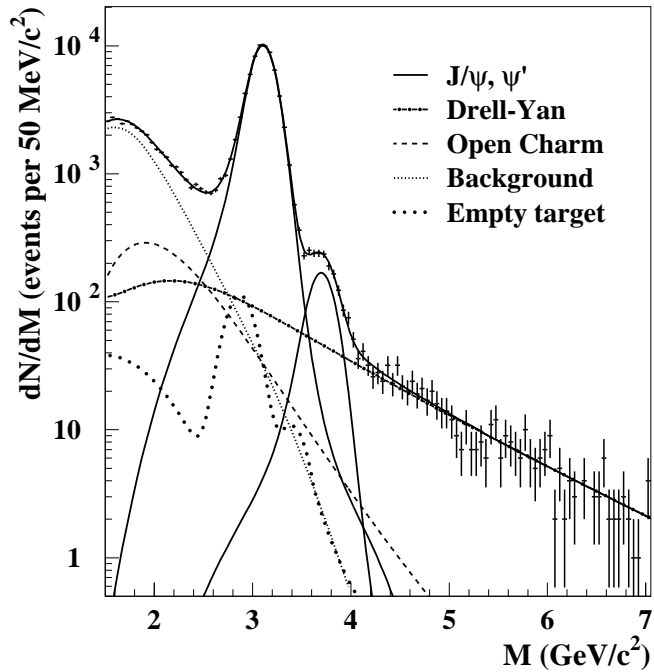}}
\caption{Measured opposite-sign dimuon mass spectrum for the p-Pb data
sample, compared with the superposition of expected contributions.}
\label{ppbfit}
\end{figure}

For dimuon masses above 2.7~GeV/$c^2$, the combinatorial background
contributes less than 1\,\% of the total number of events.  In this
mass region the relevant contributions are Drell-Yan dimuons and the
\jpsi\ and \psip\ resonances.  We have also taken into account the
contribution from semimuonic decays of D meson pairs, although it is
significantly smaller.  Figure~\ref{ppbfit} shows the opposite-sign
muon pair distribution collected in p-Pb collisions, compared to the
superposition of the different contributions, with yields obtained
through a multi-step fit procedure, explained below.

The \jpsi, \psip, Drell-Yan and open charm contribution shapes have 
been obtained by Monte-Carlo simulation, using a detailed description 
of the experimental setup and of the effects induced by multiple
scattering, energy loss, trigger requirements, etc.  The simulated 
events were reconstructed with the same programs as the real data and 
had to survive the same selection criteria.

The Drell-Yan and open charm contributions were generated with the
Monte-Carlo event generator PYTHIA~\cite{Pythia} (version 6.125) using
the GRV 94 LO set of parton distribution functions~\cite{GRV94LO} and
a primordial $k_{\mathrm{T}}$ Gaussian distribution, with width 0.8~GeV/$c$
and 1.0~GeV/$c$ for Drell-Yan and open charm, respectively~\cite{capelliPhD}.

\begin{figure}[ht]
\centering
\begin{tabular}{cc}
\resizebox{0.48\textwidth}{!}{%
\includegraphics*{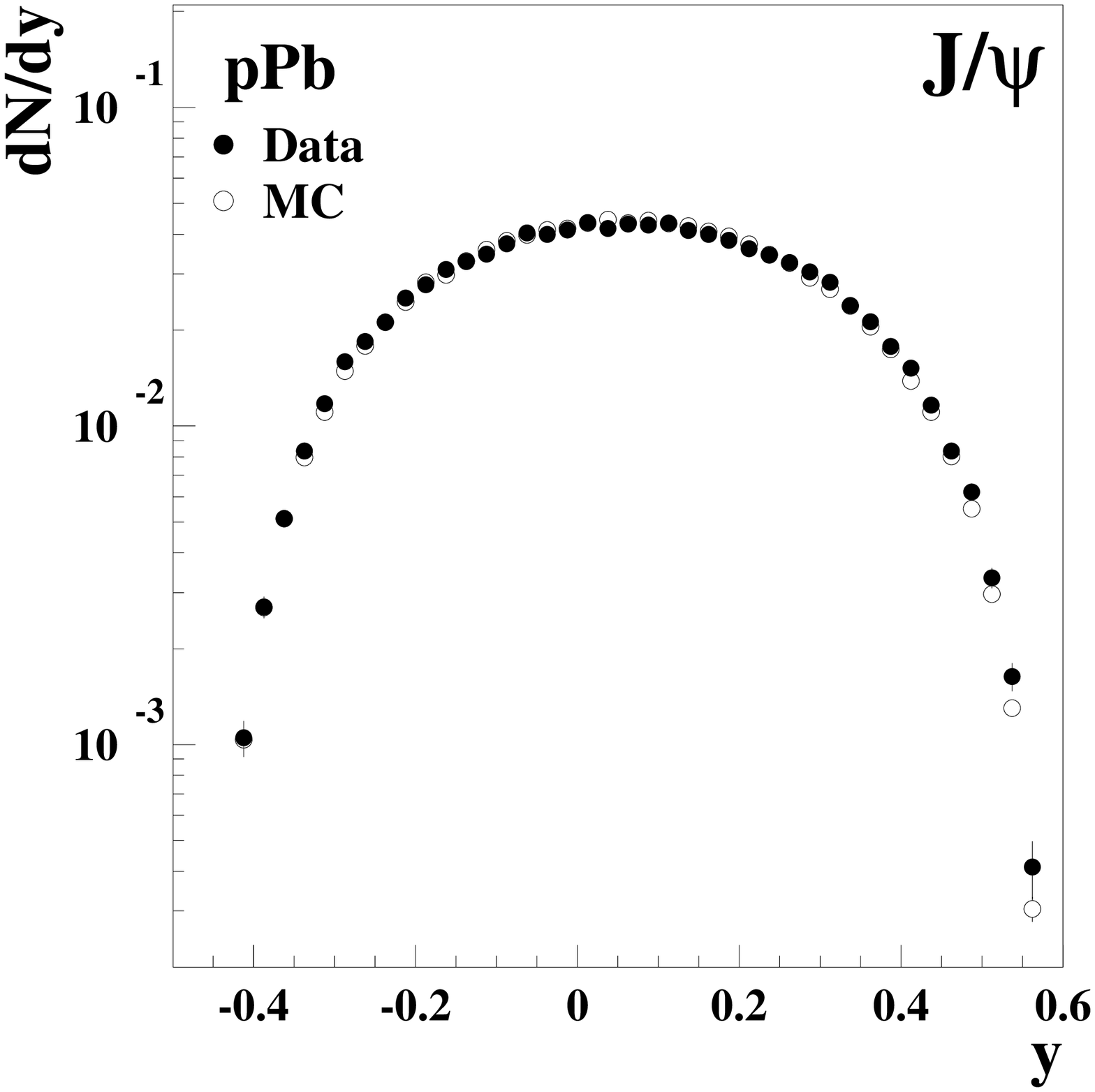}}
&
\resizebox{0.48\textwidth}{!}{%
\includegraphics*{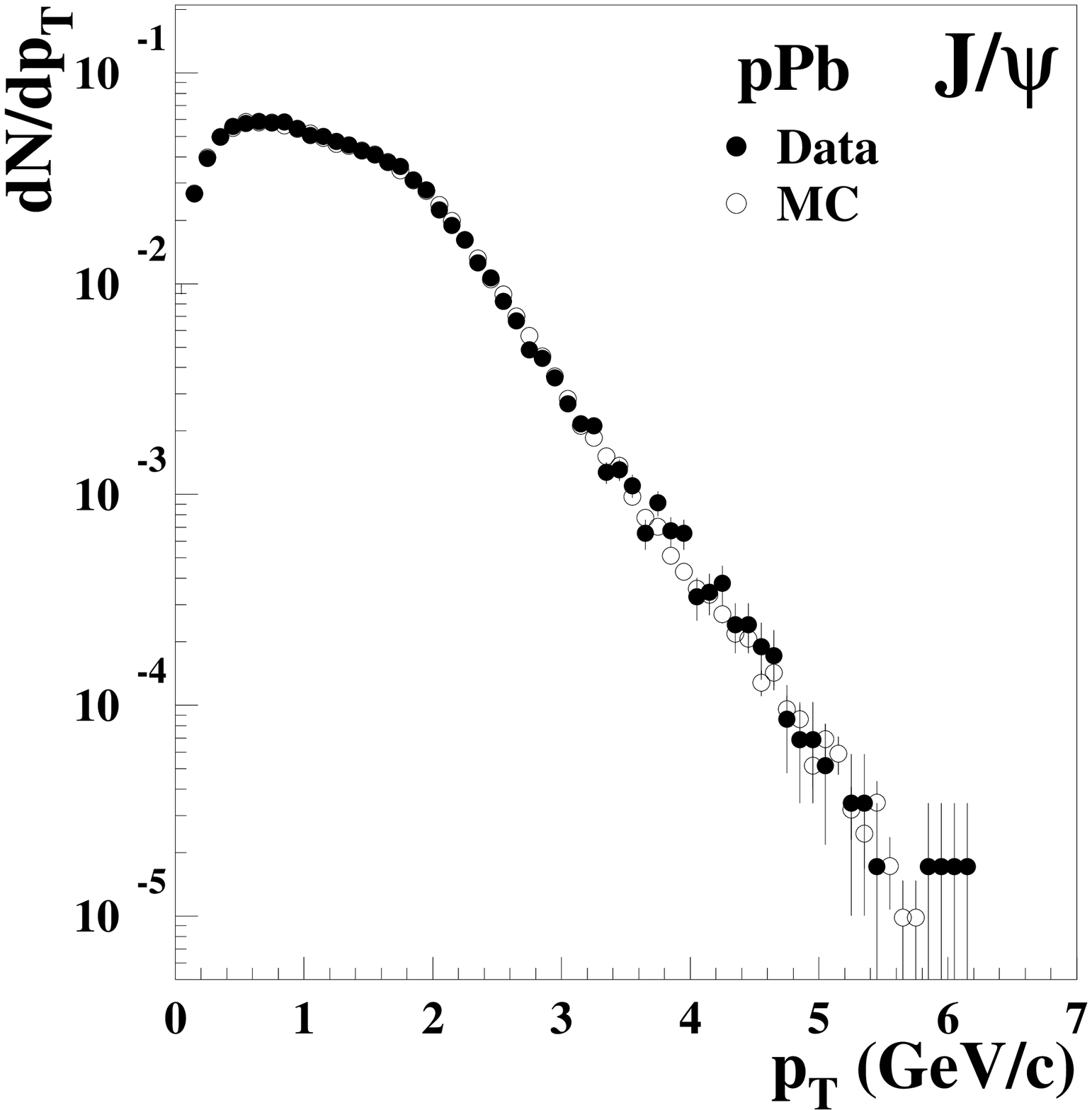}}
\end{tabular}
\caption{Measured and simulated $y$ (left) and \pt\ (right)
distributions for \jpsi\ events.}
\label{jppty}
\end{figure}

The \jpsi\ and \psip\ events were generated with a uniform
$\cos \theta_{\rm CS}$ distribution and with Gaussian rapidity
distributions, of mean $y_0=-0.2$ and width $\sigma=0.85$, common to
all targets.  These numerical values were obtained from the analysis
of proton-nucleus data samples previously collected by NA50, at
450~GeV, with about 10--15 times higher statistics (the ``HI~96/98''
data sets, described in detail in Section~6). The $p_{\rm T}$
distributions were generated according to a superposition of a
thermal function in transverse mass, $m_{\rm T} K_1(m_{\rm T}/T)$
(good for low $p_{\rm T}$), with a harder component, $\left[1+\left(
p_{\rm T}/p_0 \right)^2 \right]^{-6}$, to describe the high $p_{\rm
T}$ tail.  The measured $p_{\rm T}$ distributions were well described
with $T$ values ranging from 266~MeV (for Be) to 288~MeV (for Pb) and
a common $p_0=2.349$~GeV/$c$ parameter.  Figure~\ref{jppty} compares
the measured and simulated $y$ and \pt\ distributions, for dimuons in
the mass region \mbox{$2.9<M<3.3$~GeV/$c^2$}, where the \jpsi\ 
contributes more than 95\,\% of all events. 

The dimuon mass distributions of the \jpsi\ and \psip\ resonances, 
obtained separately for each target, are completely determined 
by experimental effects (multiple scattering and energy loss).
Their mass shapes are extracted from a fit to the 
reconstructed Monte-Carlo events using a pseudo-Gaussian 
parameterization, i.e., a Gaussian function with a mass-dependent 
width which allows to properly describe the tails of the distorted peak.
Given the very high statistics of the measured samples in the mass window 
of the resonances, even a small discrepancy between the simulated 
shapes and the measured data would significantly increase the fit 
$\chi^2$ and, thus, affect the fitting results.
Therefore, the parameters of the function describing the 
widths and tails of the charmonia mass distributions are eventually left 
free and adjust on the observed real data (see details in Ref.~\cite{GborgesPhD}).

The Monte-Carlo simulation also provides the detection acceptances of
each relevant physics process, as the ratio between reconstructed and
generated events, in the kinematical domain mentioned above, which 
amount to 13.5\,\%, 16.2\,\% and 14.4\,\%, respectively for the \jpsi, 
\psip\ and Drell-Yan dimuons (in the mass window 2.9--4.5~GeV/$c^2$).
The Monte Carlo generation functions used to compute the 
charmonia acceptances were tuned in such a way that they describe 
the experimental data (see again Fig.~\ref{jppty}). Since our final 
results have shown to be essentially insensitive to reasonable small 
changes of the Monte-Carlo distributions, no other systematic uncertainty 
has been assigned to the acceptance calculation apart from the one 
arising from the finite generated sample, reported in Table~\ref{syserr}. 
The Drell-Yan acceptances change by $\sim$0.5--1.5\% in the 
2.9-4.5\,GeV/c$^2$ mass region, systematically for all targets, when 
computed using PYTHIA and different parameterizations of parton distribution 
functions.

The opposite-sign dimuon mass spectra were fitted, independently for
each data sample, to the superposition of the contributions previously
mentioned, in the mass region between 1.5 and 8.0~GeV/$c^2$, using the 
maximum likelihood method. The combinatorial background and the empty 
target components were kept fixed, while the normalisations of the 
``signal contributions'' (\jpsi, \psip, Drell-Yan and open charm) were 
left as free parameters. 
The systematic uncertainties in the extraction of the number of 
events were investigated fitting the mass spectra under different approaches, 
such as various descriptions of the mass range below 2.3~GeV/$c^2$, changing 
the final fit mass range, checking the influence of reasonable parameterizations 
of \jpsi\ tails under the continuum, using different parton distribution functions 
for Drell-Yan generation, etc. From the previous study, we were able to select
the best inputs for the fit procedure based on the stability of the results as well as 
on a $\chi^2$/ndf criterium. However, the $\chi^2$/ndf insensitiveness to the used 
parton distribution functions and to reasonable different tails of the \jpsi\ resonance
led to somewhat arbitrary choices. As a consequence, systematic uncertainties due to 
these sources must be assigned to the measured number of events, as described in 
Section~4 and numerically presented in Table~\ref{syserr}. The final adopted analysis 
procedure provides the best overall description of the experimental data. 
Table~\ref{datafits} reports the number of \jpsi, \psip\ and Drell-Yan events (with 
their statistical error) fitted from each data set, together with the corresponding 
$\chi^2$/ndf values showing the good quality of the fits.

\begin{table}[ht]
\begin{center}
\catcode`?=\active \def?{\kern\digitwidth}
\begin{tabular}{lcccc}
\hline
\rule{0pt}{0.45cm}
Data set & $N_{\rm J/\psi}$ & $N_{\psi^\prime}$ & $N^{2.9-4.5}_{\rm DY}$ &
 $\chi^2$/ndf \\ \hline
 p-Be & 37509 $\pm$ 204 &  ?786 $\pm$ 38 & ?756 $\pm$ 43 & 0.75 \\
 p-Al & 47341 $\pm$ 229 &  1070 $\pm$ 44 & ?999 $\pm$ 50 & 1.02 \\
 p-Cu & 44224 $\pm$ 220 &  ?849 $\pm$ 40 & 1002 $\pm$ 49 & 1.18 \\
 p-Ag & 40738 $\pm$ 212 &  ?784 $\pm$ 40 & 1000 $\pm$ 49 & 1.19 \\
 p-W  & 48570 $\pm$ 232 &  ?835 $\pm$ 43 & 1233 $\pm$ 56 & 1.11 \\
 p-Pb & 67958 $\pm$ 275 &  1200 $\pm$ 51 & 1644 $\pm$ 64 & 1.01 \\
\hline
\end{tabular}
\end{center}
\vglue-2mm
\caption{Number of \jpsi, \psip\ and Drell-Yan events, 
with their statistical error, and reduced $\chi^2$ of each data 
sample fit.}
\label{datafits}
\end{table}

\section{Cross-sections results}

The production cross-sections are derived from the fitted number of
events, the integrated luminosity, the detection acceptances and from  
the different efficiencies affecting our measurement and analysis
procedure, already discussed and presented in the previous sections.

Table~\ref{abs-xsec} shows the resulting \jpsi, \psip\ and Drell-Yan
production cross-sections, per target nucleon, in the phase space window 
of our measurement.  The values for charmonia include the branching
ratios of their decay into muons. 
The Drell-Yan production cross-sections 
presented in this table (given for the mass range 2.9--4.5~GeV/$c^2$) have 
been multiplied by ``isospin correction factors'' \cite{pcpaper} between 
0.954 and 0.947. The isospin corrected cross-sections are equivalents to 
the ones that would have been measured if the target nuclei were made of
protons only.

\begin{table}[ht]
\centering
\begin{tabular}{lccc}
\hline
    & $B_{\mu\mu}\sigma(\rm J/\psi)$ / $A$ 
    & $B^\prime_{\mu\mu}\sigma(\psi^\prime)$ / $A$ 
    & $\sigma(\rm DY_{2.9-4.5})$ / $A$ \\
    & (nb/nucleon) & (pb/nucleon) & (pb/nucleon) \\ \hline
 p-Be & 4.717 $\pm$ 0.026 (2.1\%) & 82.3 $\pm$ 4.0 (2.1\%) & 86.4 $\pm$ 5.0 (2.1\%)\\
 p-Al & 4.417 $\pm$ 0.022 (2.1\%) & 83.5 $\pm$ 3.4 (2.1\%) & 84.2 $\pm$ 4.4 (2.1\%)\\
 p-Cu & 4.280 $\pm$ 0.022 (2.1\%) & 68.2 $\pm$ 3.2 (2.7\%) & 86.7 $\pm$ 4.3 (2.2\%)\\
 p-Ag & 3.994 $\pm$ 0.022 (2.1\%) & 63.9 $\pm$ 3.3 (2.1\%) & 86.9 $\pm$ 4.4 (2.1\%)\\
 p-W  & 3.791 $\pm$ 0.019 (2.1\%) & 53.9 $\pm$ 2.9 (2.5\%) & 84.4 $\pm$ 3.9 (2.1\%)\\
 p-Pb & 3.715 $\pm$ 0.016 (2.1\%) & 54.3 $\pm$ 2.4 (2.6\%) & 79.1 $\pm$ 3.2 (2.2\%)\\
\hline
\end{tabular}
\vglue-2mm
\caption{Production cross-sections, per target nucleon, for \jpsi\ and
\psip\ production (times branching ratios into muons), and for Drell-Yan 
dimuons in the mass range 2.9--4.5~GeV/$c^2$, integrated in the phase 
space window of our measurement ($-0.425< y_{\rm cm} <0.575$ and 
$-0.5< \cos \theta_{\rm CS} <0.5$).  The values in parentheses
indicate the relative systematic uncertainties which are not common to
all targets.}
\label{abs-xsec}
\end{table}

Since the data were collected with relatively frequent target changes, 
most time-dependent systematic effects (such as the ones affecting the 
beam intensity measurement) should be identical among the six data sets.  
With respect to the data samples previously collected by NA50, when each 
p-A system was studied in a different data taking period (often meaning 
different years), this running mode significantly reduces the systematic 
uncertainty in the determination of the nuclear dependences, where only 
the relative changes of the production yields, from target to target, 
are important. This is the reason why the systematic errors quoted in
Table~\ref{abs-xsec} only refer to uncertainties specific to each data
set. These uncertainties may be separated in two different 
kinds of sources: the ones originated by the experimental setup performance
($\sim$\,0.4--0.7\,\% from the target material properties; $\sim$\,0.2\,\% 
from the reconstruction efficiencies; $\sim$\,2.3\,\% from the dimuon trigger 
efficiency), and the ones related to the extraction of the physical yields, as 
described in the Section~3 (0.2\,\% ``resp. 0.4\,\%'' for the charmonia 
``resp. Drell-Yan'' acceptance and $\sim$\,0.4--1.8\,\% ``resp. $<$1\,\%'' for 
\psip\ ``resp. Drell-Yan'' due to the \jpsi\ high mass tail uncertainty). 
Still concerning the Drell-Yan yields, their low statistics do not allow us 
to choose which set of parton distribution functions better describes the 
experimental results. This gives rise to a -0.3\% to +2.8\,\% uncertainty 
(including acceptance corrections), to be considered when comparing our results 
with those obtained from other data sets, and which affects systematically changes 
the results of each and every target in the same way.

Finally, the luminosity was computed from the integrated number of 
incident protons within the spectrometer lifetime (see Section~2 for details) 
and from the effective number of target particles per unit surface. The 
uncertainty on the number of incoming protons was estimated to be 3\,\%  
based on the systematic error of the latest calibration of the ionization 
chambers~\cite{GborgesPhD}.
This value also affects all targets in the same way and was not included in 
Table~\ref{abs-xsec}. However, it must be taken into account when 
comparing these results with those obtained from data collected in 
previous years, in what concerns their absolute normalisation.

All the different sources specific to each of the targets are summarized 
in Table~\ref{syserr} and a detailed discussion can be found in 
Ref.~\cite{GborgesPhD}.

\begin{table}[ht]
\centering
\begin{tabular}{lcccccc}
\hline
Sources                                   & p-Be      & p-Al    & p-Cu     & p-Ag     & p-W      & p-Pb    \\ \hline
Experimental setup uncertainties          &           &         &          &          &          &         \\ 
$\epsilon_{\rm trig}$ (\%)                & 2.3       & 2.3     & 2.3      & 2.3      & 2.3      & 2.3     \\
$\epsilon_{\mu\mu}$ (\%)                  & 0.2       & 0.2     & 0.2      & 0.2      & 0.2      & 0.1 	   \\ 
$N_{\rm tgt}$ (\%)                  	  & 0.5       & 0.4 	& 0.4 	   & 0.4      & 0.5 	 & 0.7 	   \\ \hline
Yields uncertainties                      &           &         &          &          &          &         \\ 
$\cal A$(\jpsi, \psip) (\%)               & 0.2       & 0.2 	& 0.2 	   & 0.2      & 0.2 	 & 0.2 	   \\
$\cal A$(DY$_{2.9-4.5}$) (\%)             & 0.4       & 0.4 	& 0.4      & 0.4      & 0.4 	 & 0.4 	   \\ 
\jpsi~high tail; \jpsi\ (\%)	          & $<$ 0.1   & $<$ 0.1	& $<$ 0.1  & $<$ 0.1  & $<$ 0.1	 & $<$ 0.1 \\
\jpsi~high tail; \psip\ (\%)	          & 0.4       & 0.5 	& 1.8 	   & 0.5      & 1.4 	 & 1.5 	   \\
\jpsi~high tail; DY$_{2.9-4.5}$ (\%)      & 0.2       & 0.3     & 0.8 	   & 0.2      & 0.4      & 0.4 	   \\ \hline
\end{tabular}
\vglue-2mm
\caption{Contributions to the overall systematic uncertainty 
of cross-section measurements, specific to each of the targets.}
\label{syserr}
\end{table}

Table~\ref{rat-xsec} presents the ratios between the \jpsi, \psip\ and
Drell-Yan production cross-sections.  Their uncertainties are dominated 
by the statistical errors of the Drell-Yan and \psip\ values.


\begin{table}[ht]
\begin{center}
\begin{tabular}{lccc}
\hline
    &  $\frac{\rm B_{\mu\mu}\sigma(J/\psi)}{\rm \sigma(DY_{2.9-4.5})}$ 
    &  $\frac{\rm B^\prime_{\mu\mu}\sigma(\psi^\prime)}{\rm \sigma(DY_{2.9-4.5})}$ 
    &  \rule{0pt}{0.45cm} $\frac{\rm B^\prime_{\mu\mu}\sigma(\psi^\prime)}{\rm B_{\mu\mu}
  \sigma(\psi)}~(\%)$ \\ \hline
 p-Be & 54.6 $\pm$ 3.2 (0.5\%)  & 0.953 $\pm$ 0.082 (0.5\%) & 1.745 $\pm$ 0.085 (0.6\%) \\
 p-Al & 52.5 $\pm$ 2.8 (0.6\%)  & 0.991 $\pm$ 0.076 (0.5\%) & 1.889 $\pm$ 0.078 (0.7\%) \\
 p-Cu & 49.3 $\pm$ 2.5 (0.9\%)  & 0.786 $\pm$ 0.063 (1.1\%) & 1.593 $\pm$ 0.076 (1.9\%) \\
 p-Ag & 46.0 $\pm$ 2.4 (0.5\%)  & 0.735 $\pm$ 0.061 (0.6\%) & 1.599 $\pm$ 0.084 (0.7\%) \\
 p-W  & 44.9 $\pm$ 2.1 (0.6\%)  & 0.639 $\pm$ 0.052 (1.1\%) & 1.422 $\pm$ 0.076 (1.5\%) \\
 p-Pb & 47.0 $\pm$ 1.9 (0.7\%)  & 0.687 $\pm$ 0.047 (1.2\%) & 1.461 $\pm$ 0.064 (1.6\%) \\
\hline
\end{tabular}
\end{center}
\vglue-2mm
\caption{Ratios of \jpsi, \psip\ and Drell-Yan production cross-sections, 
in our phase space window ($-0.425< y_{\rm cm} <0.575$ and 
$-0.5< \cos \theta_{\rm CS} <0.5$). The values in parentheses
indicate the relative systematic uncertainties.} 
\label{rat-xsec}
\end{table}

\section{Nuclear dependences}

The main topic of this work is the study of the nuclear dependence of
the \jpsi\ and \psip\ production cross-sections.  It is generally
accepted that charmonia \emph{production}, essentially due to gluon
fusion, is a hard process and, therefore, is basically independent of
the amount of nuclear matter surrounding the production point.  This
is also the case of high mass dimuon (Drell-Yan) production, where a
linear scaling of the production cross-section with the target nucleus
mass number, in p-A collisions, has been observed~\cite{Ald91,Ald90}.
This linearity is usually expressed by saying that $\alpha$ is unity,
where $\alpha$ is the exponent in the power law $\sigma_{pA}=\sigma_0
A^{\alpha}$ (the ``$\alpha$ parameterization'').  It has been observed
that $\alpha$ decreases as a function of \xf\ and increases as a
function of \pt\ (the ``Cronin effect'').  Since these changes affect
Drell-Yan production (where no final state effects are expected), they
are attributed to initial state effects: nuclear modifications of the
parton distribution functions (shadowing, etc.) and energy loss of the
incoming partons as they traverse the target nucleus~\cite{Vogt}.
Charmonia production, besides these initial state effects, is also
sensitive to nuclear absorption of the produced \ccbar\ bound state,
before and after forming the final \jpsi\ or \psip\ resonances,
through strong interactions with the nuclear environment on their way
out of the target nucleus.

\begin{figure}[ht]
\centering
\begin{tabular}{cc}
\resizebox{0.48\textwidth}{!}{%
\includegraphics*{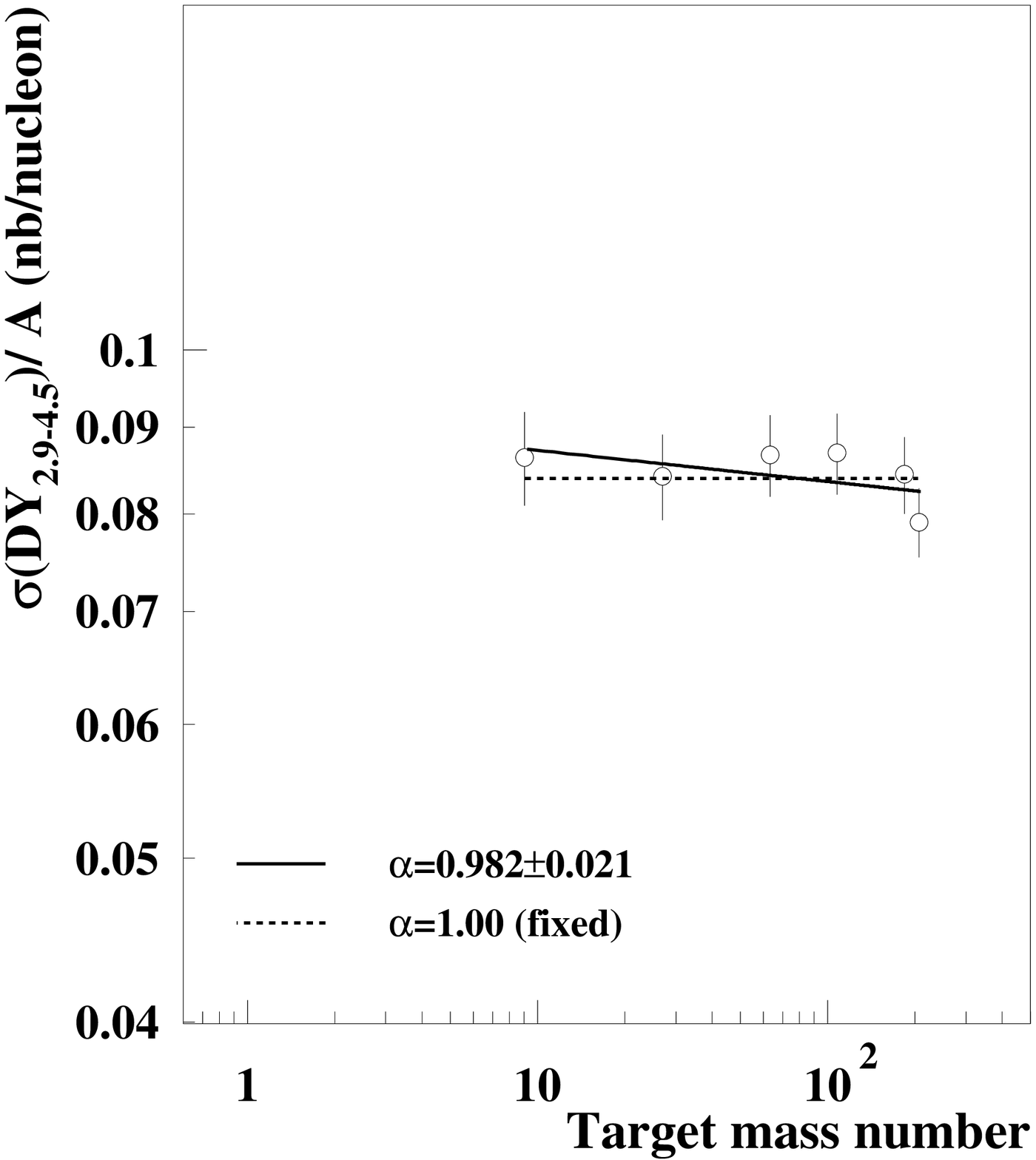}}
&
\resizebox{0.48\textwidth}{!}{%
\includegraphics*{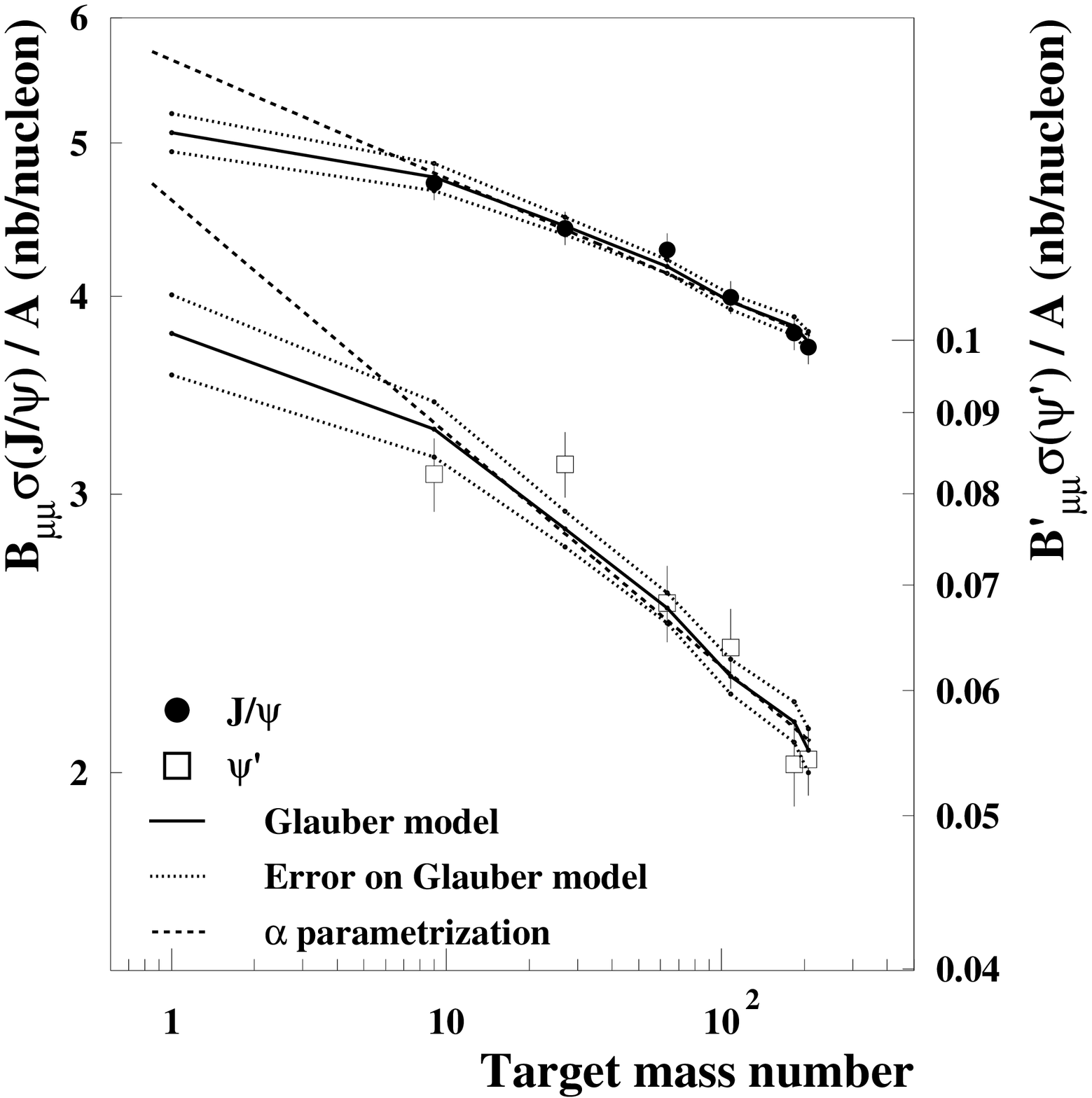}}
\end{tabular}
\caption{Drell-Yan (left) and charmonia (right) production
cross-sections, per nucleon, as a function of the target mass number.
Lines are fits to the $\alpha$ parameterization or to the Glauber
model, as indicated.}
\label{alphfit}
\end{figure}

The left panel of Fig.~\ref{alphfit} shows the Drell-Yan production
cross-section, per target nucleon, as a function of the target mass
number, as extracted from our six data samples.  The flat pattern
confirms the linear scaling previously observed in several proton-nucleus 
experiments~\cite{Ald91,Ald90}, including NA50~\cite{pcpaper}, indicating 
that \emph{initial} state effects play a negligible role in Drell-Yan 
production at our energies and in the NA50 phase space window (mid-rapidity). 
From the present analysis we extract $\alpha_{\rm DY}=0.982\pm0.021$.  
The right panel of Fig.~\ref{alphfit} shows the corresponding results 
for the \jpsi\ and \psip\ production cross-sections, which lead to
$\alpha$ values smaller than unity: $\alpha_{\rm J/\psi}=
0.925\pm0.009$ and $\alpha_{\psi^\prime}=0.852\pm0.019$.

A much more accurate formalism to describe the observed charmonium
production cross-sections is provided by the Glauber 
model~\cite{glauber}.  Within this framework, a proton-nucleus or
nucleus-nucleus interaction is considered as a set of independent
interactions between the projectile and target nucleons, assuming that
the properties of the nucleons do not change after the first collision
and that they can interact further with the same cross-section.  The
basic input for the Glauber formalism in \mbox{p-A} collisions is the
target nuclear thickness, $T_A(\vec{b})$, representing the number of
nucleons per unit surface at impact parameter $b$.  The nuclear
density profiles of light nuclei ($A<17$) were parameterized using the
Fermi oscillator model.  For the other nuclei we used Woods-Saxon
distributions, with the parameters given in~\cite{Jag}, except for the
Ag nucleus, missing in those tables, for which we have used the
parameterization of the FRITIOF~7.02 Monte Carlo code~\cite{And87}.
The data were then fitted with the expression
$$\frac{\sigma^\psi}{A}=\frac{\sigma_{\rm 0}^{\rm G}}{ (A-1)~
\sigma^{\rm G}_{\rm abs}} \int {\rm d}\vec{b} ~(1-\exp(-(A-1)
\, \sigma^{\rm G}_{\rm abs}\, T_{A}(\vec{b}))) \quad,$$ leaving as free
parameters the normalisations, $\sigma_{\rm 0}^{\rm G}$, and the
interaction cross-sections of the \jpsi\ or $\psi^\prime$ on their way
through the nuclear matter, $\sigma^{\rm G}_{\rm abs}$.

\begin{figure}[ht!]
\centering
\resizebox{0.6\textwidth}{!}{%
\includegraphics*{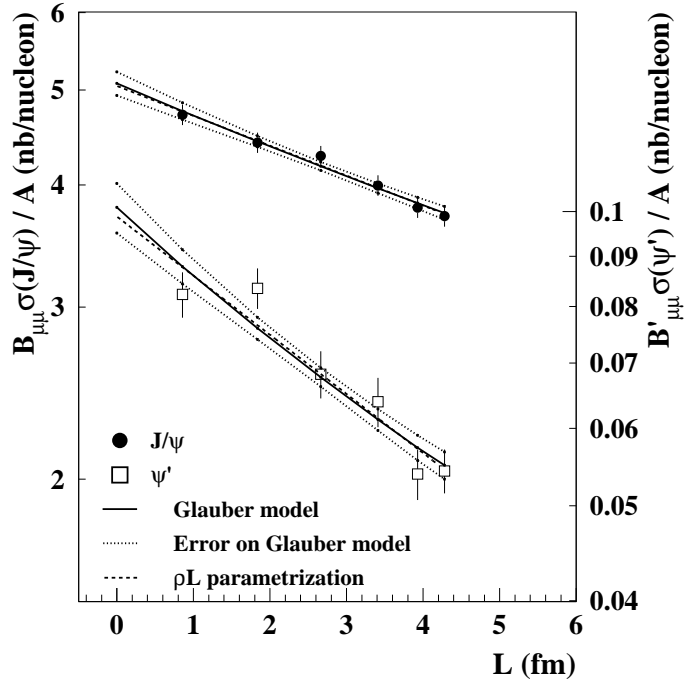}}
\caption{Measured \jpsi\ and \psip\ production cross-sections, per
target nucleon, fitted with the full Glauber model and with the $\rho
L$ approximation.}
\label{rholglbfit}
\end{figure}

The resulting fits to the \jpsi\ and \psip\ production cross-sections
are presented in Figs.~\ref{alphfit} (right) and~\ref{rholglbfit},
respectively as a function of $A$, the target mass number, and $L$,
the average distance of nuclear matter crossed by the charmonium
states on their way out of the nucleus.  The solid lines join together
the results of the Glauber calculations, made independently for each
individual target, while the dotted lines represent the error bands
resulting from the uncertainties on the normalisations and absorption
cross-sections.

The detailed Glaubel model formalism can be approximated, in first
order, by the ``$\rho L$ parameterization'',
$$\sigma_{pA} = \sigma_0^{\rho L} \, A \, \exp(-\sigma^{\rho L}_{\rm abs}
\cdot \langle {\rho L} \rangle) \quad,$$ where $\sigma^{\rho L}_{\rm
abs}$ is the (approximate) absorption cross-section of the charmonia
state and $\rho$ the density of the nuclear matter.  $L$ is computed,
for each target nucleus, in the framework of the Glauber model, as
explained in~\cite{rspaper}.  The corresponding fits, shown in
Fig.~\ref{rholglbfit} as dashed lines, are almost indistinguishable
from the full Glauber calculations, contrary to what happens with the
$\alpha$ parameterization, which significantly diverges from the
correct calculation in the region of very light nuclei, as shown in
Fig.~\ref{alphfit} (right) and discussed in detail in~\cite{rspaper,RubenPhD}.
The results of the fits are collected in Table~\ref{fitres}.

\begin{table}[ht!]
\begin{center}
\catcode`?=\active \def?{\kern\digitwidth}
\begin{tabular}{c|ccc}
\hline
 $\alpha$ fit &  $\sigma_0^\alpha$ &  $\alpha$ &  $\chi^2$/ndf \\ \hline
 \jpsi &  ?5.6 $\pm$ 0.2 nb/nuc & $0.925\pm0.009$ & 0.8 \\ 
 \psip &  123 $\pm$q0 pb/nuc & $0.852\pm0.019$ & 2.0 \\ \hline
\multicolumn{4}{c}{} \\ \hline
\rule{0pt}{0.45cm}
 Glauber fit &  $\sigma_0^{\rm G}$ &  $\sigma^{\rm G}_{\rm abs}$ (mb) &  $\chi^2$/ndf \\ \hline
 \jpsi & ?5.1 $\pm$ 0.1 nb/nuc & ?4.6 $\pm$ 0.6     & 0.4 \\ 
 \psip & 101 $\pm$ 6  pb/nuc  & 10.1 $\pm$ 1.6    & 1.7 \\ \hline
\multicolumn{4}{c}{} \\ \hline
\rule{0pt}{0.45cm}
 $\rho L$ fit &  $\sigma_0^{\rho L}$ &  $\sigma^{\rho L}_{\rm abs}$ (mb) &  $\chi^2$/ndf \\ \hline
 \jpsi & ?5.0 $\pm$ 0.1 nb/nuc & 4.1 $\pm$ 0.5     & 0.3 \\  
 \psip &  99 $\pm$ 5  pb/nuc  & 8.2$\pm$1.0     & 1.5 \\  \hline
\end{tabular}
\end{center}
\vglue-2mm
\caption{Charmonia nuclear dependence parameters, extracted with the
$\alpha$-parameterization, the Glauber model, and its $\rho L$
approximation.}
\label{fitres}
\end{table}

\begin{figure}[ht]
\centering
\begin{tabular}{cc}
\resizebox{0.48\textwidth}{!}{%
\includegraphics*{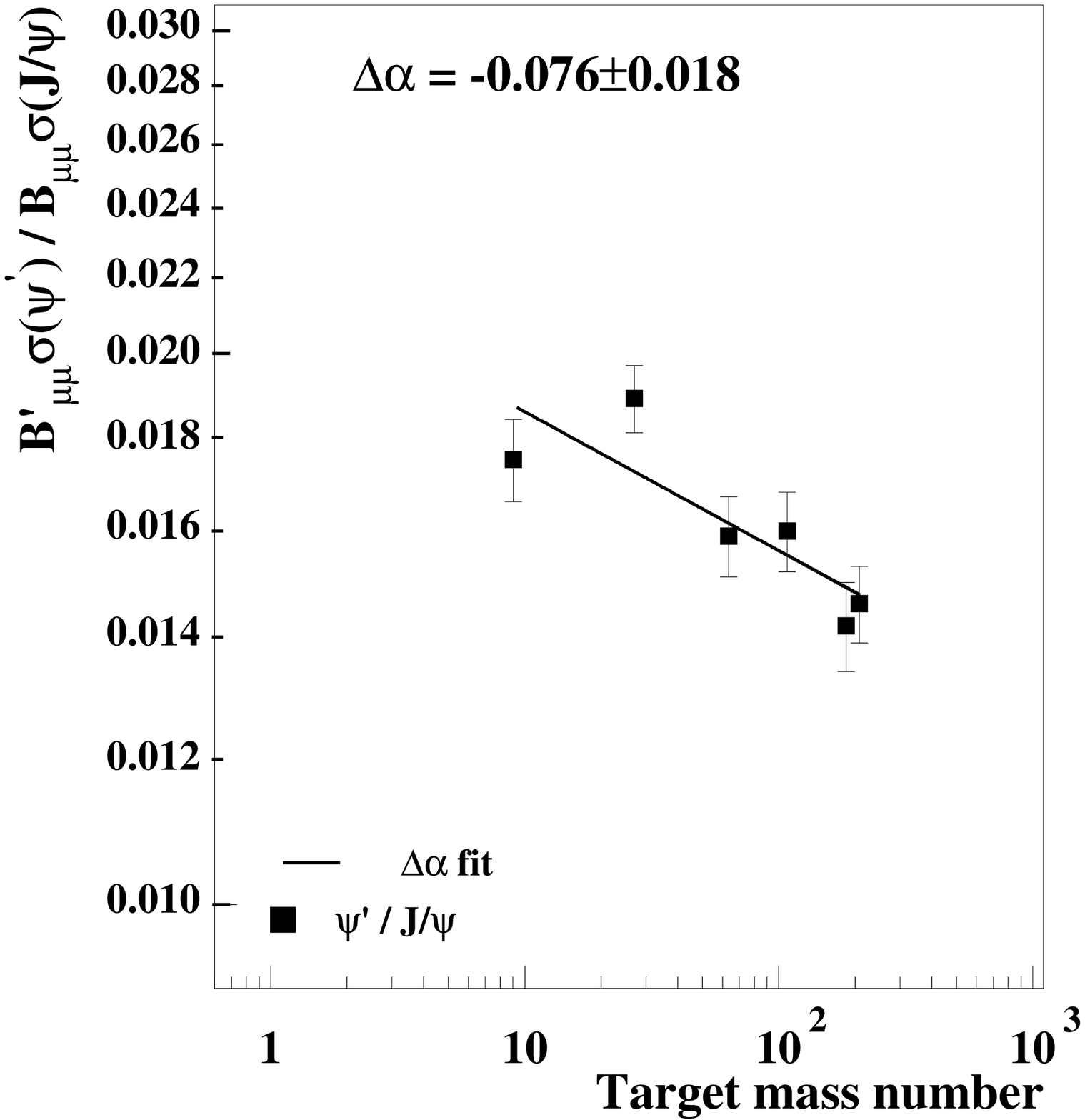}}
&
\resizebox{0.48\textwidth}{!}{%
\includegraphics*{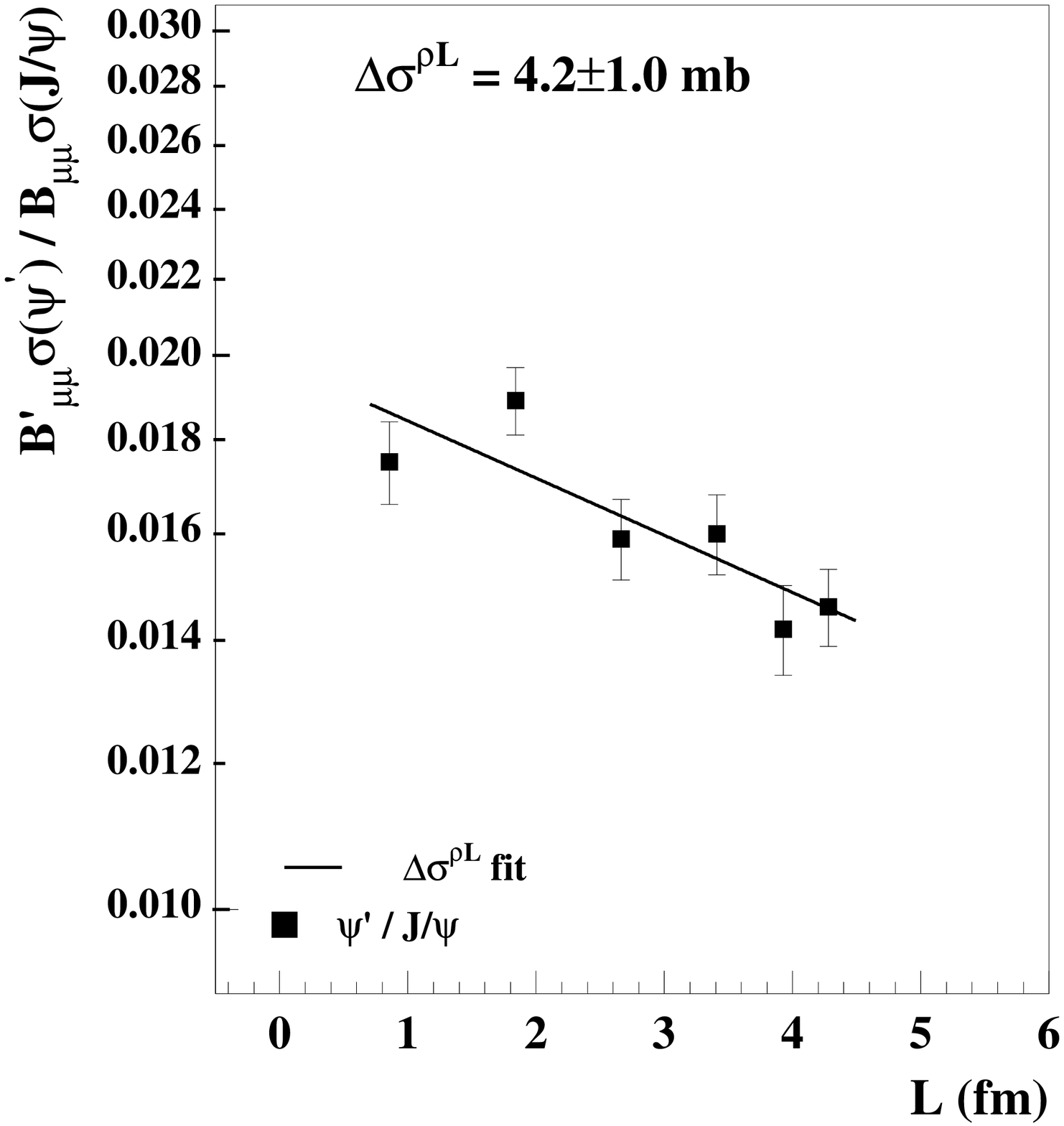}}
\end{tabular}
\caption{Ratio between the \psip\ and \jpsi\ production
cross-sections, as a function of $A$ (left) and $L$ (right).}
\label{alprholfit}
\end{figure}

We see that the \psip\ suffers a stronger nuclear absorption than the
\jpsi.  However, the relative differences between the nuclear
dependences of the two charmonium states can be studied with better
accuracy through the ratio $B^\prime_{\mu\mu}\sigma(\psi^\prime) /
B_{\mu\mu}\sigma(\rm J/\psi)$, shown in Fig.~\ref{alprholfit} as a
function of $A$ (left) and $L$ (right).  This ratio cancels out part
of the systematic uncertainties, which affect both charmonium states
in the same way.

Fitting these measurements as a function of the target mass number,
with the $\alpha$ parameterization, or as a function of $L$, with the
$\rho L$ parameterization, we can directly obtain the parameters
which characterize the \emph{relative} nuclear dependences: \mbox{$\Delta \alpha
= \alpha(\psi^\prime) - \alpha(\rm J/\psi)$} and \mbox{$\Delta \sigma^{\rho
L}_{\rm abs} = \sigma^{\rho L}_{\rm abs}(\psi^\prime) - \sigma^{\rho
L}_{\rm abs}(\rm J/\psi)$}.  It is not possible to directly extract
$\Delta \sigma^{\rm G}_{\rm abs}$ with a Glauber fit.  The
corresponding results are collected in Table~\ref{fitresppjp}.

\begin{table}[ht]
\begin{center}
\begin{tabular}{c|ccc}
\hline
$\alpha$ fit   &  $\sigma_0^\alpha(\psi^\prime / {\rm
J/\psi})$ (\%) &  $\Delta \alpha$ &  $\chi^2$/ndf \\ \hline 
               & $2.21\pm0.17$ & $-0.076\pm0.018$ & 1.9 \\ \hline 
\multicolumn{4}{c}{} \\ \hline
\rule{0pt}{0.45cm}
$\rho L$ fit   &  $\sigma_0^{\rho L}(\psi^\prime/{\rm
J/\psi})$ (\%) &  $\Delta \sigma^{\rho L}_{\rm abs}$ (mb) &  $\chi^2$/ndf \\ \hline 
               & $1.98\pm0.10$ & $4.2\pm1.0$      & 1.7 \\ \hline
\end{tabular}
\end{center}
\vglue-2mm
\caption{Relative absorption measured directly from the
${\rm B^\prime_{\mu\mu}\sigma(\psi^\prime)}/{\rm B_{\mu\mu}
\sigma(\psi)}$ ratios, and fitted with the $\alpha$ and $\rho L$
parameterizations.}
\label{fitresppjp}
\end{table}

The nuclear absorption of the charmonium states can also be studied through the 
ratios of their production cross-sections with respect to the Drell-Yan cross-section, 
\jpsi\,/\,DY and \psip\,/\,DY.  Under the assumption (supported by existing results
\cite{Ald91,pcpaper}) that Drell-Yan production has negligible initial and 
final state effects, the values extracted from these ratios should be essentially 
identical to those obtained from the \jpsi\ and \psip\ absolute production cross-sections.  
The rather low Drell-Yan statistics of the data samples collected in year 2000, however, 
do not allow us to make significant comparisons between the two approaches.  For instance, 
the Glauber formalism applied to the \jpsi\,/\,DY ratios gives 
$\sigma^{\rm G}_{\rm abs}(\rm J/\psi) = 3.4\pm1.2$~mb, while the absolute \jpsi\ 
cross-section analysis gives $4.6\pm0.6$~mb.  Nevertheless, for completeness, we give the 
values extracted from the analysis of the \jpsi\,/\,DY and \psip\,/\,DY ratios at the end of 
the next section (Table~\ref{tab:ratiossim}), together with the values extracted from the 
simultaneous analysis of all NA50 p-A data sets.

\section{Comparison with previous NA50 p-A results}

We will now compare the results shown in the previous sections with
those obtained by NA50 using proton-nucleus data collected in previous
years.  For the purpose of this comparison, we will label the data
sets analysed in detail in the present paper as ``HI~2000'' (given the
``high intensity'' of the proton beam and the year of data taking).

NA50 collected its first p-A data sets between 1996 and 1998, with a
high intensity 450~GeV proton beam.  Five different targets were used
(Be, Al, Cu, Ag and W), each one during a period of around one week.
These data samples, referred as ``HI~96/98'', are the ones with the
highest statistics (between 10 and 15 times higher than the data sets
analysed in the present paper).  

The high intensity beam induced high hit multiplicities in the wire
chambers, which the standard offline reconstruction software used by
NA50 in the first years had difficulties in handling properly.
This motivated a significant upgrade of the reconstruction software,
after which the data could be reconstructed with higher efficiency.
The analysis details and results can be found in~\cite{rspaper,RubenPhD}.

In parallel, new p-A data samples were collected at a reduced beam
intensity ($\sim$\,1--5~$10^8$ protons per burst).  The energy of the
beam was 450~GeV and the targets used were the same as before.  These
data sets, referred hereafter as ``LI~98/00'', have 20--30\,\% of the
statistics of the previous ones.  Two independent analyses were
performed using these data, in slightly different kinematical windows,
and the results have been published in~\cite{pcpaper,rspaper}.

Given the improvements in the reconstruction software, the ``HI~2000''
data sets were collected at high beam intensities, even slightly
higher than those of the ``HI~96/98'' samples.  However, the short
beam time available prevented the collection of sizeable statistics.
On the other hand, the use of all the targets in the same data taking
period significantly contributes to a decrease in the systematic
uncertainties, when determining the nuclear dependences of the
production cross-sections.

Given the slightly different analysis procedures used, the three data
samples cannot be directly compared without first performing a few
corrections.  There are three main differences among the
``HI~96/98'', ``LI~98/00'' and ``HI~2000'' data sets or analyses.  
First, the rapidity window considered for the extraction of the 450~GeV 
results is either $-0.4<y_{\rm cm}<0.6$ or $-0.5<y_{\rm cm}<0.5$ (we 
remind that the 400\,GeV probe the range $-0.425<y_{\rm cm}<0.575$). 
Second, the Drell-Yan calculations were done with different parton 
distribution functions: MRS~G~\cite{MRS41}, MRS~A (Low Q$^2$)~\cite{MRS43}
and GRV~94~LO~\cite{GRV94LO}.  Third, the Monte Carlo simulations of
the charmonium states used somewhat different $y$ and \pt\ distributions.

For a coherent comparison of all the three data sets, we corrected 
the 450~GeV results to a common rapidity domain, $-0.5<y_{\rm cm}<0.5$,
and also so that they correspond to the conditions used in the
``HI~2000'' analysis, as described in the previous sections of this
paper: GRV~94~LO parton distribution functions and improved charmonia
rapidity and \pt\ distributions.  These corrections, described in
detail in~\cite{GborgesPhD}, are rather small: 0.9\,\% (0.2\,\%) for 
the charmonia (Drell-Yan) $y$ window, between $-2.0$\,\% and $+1.3$\,\% 
for the Drell-Yan shape imposed by the different PDFs and $\sim$\,1\,\% 
for the charmonia kinematical distributions. 
Finally, the HI~96/98 Drell-Yan results were corrected by 6.9$\pm$1.6\%,
in order to account for their systematically lower normalisation with respect
to the other data sets, as reported in \cite{RubenPhD}.

\begin{figure}[ht!]
\centering
\resizebox{0.6\textwidth}{!}{%
\includegraphics*{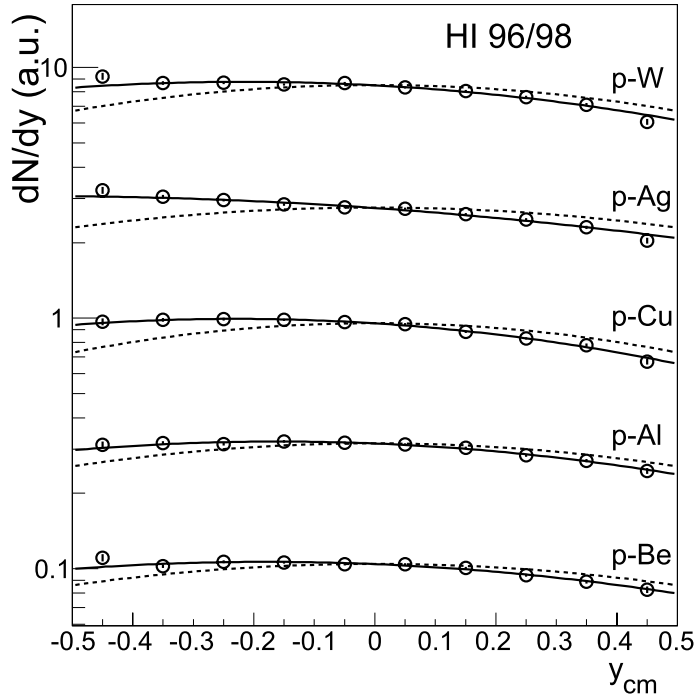}}
\caption{\jpsi\ rapidity distributions, in the centre of mass frame,
  as measured by NA50 through the ``HI~96/98'' data sets, with five
  different nuclear targets and a 450~GeV proton beam.  The data
  points are fitted to Gaussian functions with a free width and a mean
  which is either left free (solid lines) or fixed to 0 (dashed
  lines).}
\label{hi9698yfits}
\end{figure}

The change in the function used to generate the rapidity distribution
of the charmonium states is worth being explained in some detail.  The
analyses of the ``HI~96/98'' and ``LI~98/00'' data sets were done
assuming that the \jpsi\ mesons were produced with a centre of
mass rapidity distributed according to a Gaussian centred at
mid-rapidity and with a r.m.s.\ of 0.75~\cite{RubenPhD} or
0.60~\cite{PCortPhD}, independently of the target.  After these
analyses were published~\cite{pcpaper,rspaper}, we tried to understand
better the specific issue of the rapidity distributions, from a more
detailed look at our measurements, despite the fact that our rapidity
coverage is relatively narrow (only one unit in rapidity).
Figure~\ref{hi9698yfits} shows the rapidity distributions (corrected for 
acceptance) of events in the mass range $2.86<M<3.34$~GeV/$c^2$, from the 
``HI~96/98'' (highest statistics) data sample.  These distributions were 
fitted to Gaussians with free widths and either centred at mid-rapidity 
($y_0=0$, dashed lines) or with a free mean (solid lines). All the five 
different targets, from Be to W, systematically prefer a significantly 
negative mean value, $y_0\sim-0.2$, and a common width, $\sigma=0.85$, 
with reasonable fit quality, $\chi^2/{\rm ndf}\sim1$--3.  On the contrary, 
forcing the $y$ distributions to be centered at mid-rapidity leads to a 
bad description of the measured data, with $\chi^2/{\rm ndf}$ values 
between 18 and 51, depending on the target. 
The Monte Carlo simulation done in 
the present work (for the \jpsi\ and the \psip) used Gaussian rapidity distributions 
with mean $-0.2$ and $\sigma=0.85$.  Nevertheless, we have seen that the final results 
are essentially insensitive to reasonable changes in the rapidity distributions used 
for the event generation.

Table~\ref{oldpacs} compiles the ``HI~96/98'' and ``LI~98/00''
results, after the corrections just mentioned.  The ``LI~98/00''
results shown here are the average of the two independent analyses
mentioned above, since they are in agreement with each other, when
compared in a coherent way.

\begin{table}[ht]
\begin{center}
\catcode`?=\active \def?{\kern\digitwidth}
\begin{tabular}{c|cc|cc|cc}
\hline
& \multicolumn{2}{c|}{$B_{\mu\mu}\sigma(\rm J/\psi)$/A} 
& \multicolumn{2}{c|}{$B^\prime_{\mu\mu}\sigma(\psi^\prime)$/A} 
& \multicolumn{2}{c}{$\sigma(\rm DY_{2.9-4.5})$/A}\\ 
& \multicolumn{2}{c|}{(nb/nucleon)} & \multicolumn{2}{c|}{(pb/nucleon)}
& \multicolumn{2}{c}{(pb/nucleon)} \\ \hline
    & HI 96/98      & LI 98/00      & HI 96/98     & LI 98/00     & HI 96/98      & LI 98/00 \\ \hline       
 Be & 5.11$\pm$0.18 & 5.27$\pm$0.23 & 88.2$\pm$3.8 & 89.4$\pm$5.3 & ?93.3$\pm$4.1 & ?96.6$\pm$5.6 \\  
 Al & 4.88$\pm$0.23 & 5.14$\pm$0.21 & 84.3$\pm$4.6 & 88.0$\pm$4.8 & ?96.1$\pm$5.0 & 104.0$\pm$5.9 \\
 Cu & 4.74$\pm$0.18 & 4.97$\pm$0.22 & 77.7$\pm$3.4 & 82.1$\pm$4.9 & 101.3$\pm$4.6 & ?97.3$\pm$6.8 \\
 Ag & 4.45$\pm$0.15 & 4.52$\pm$0.20 & 69.8$\pm$2.7 & 72.8$\pm$4.3 & ?94.3$\pm$3.7 & 100.4$\pm$5.3 \\
 W  & 4.05$\pm$0.15 & 4.17$\pm$0.37 & 61.7$\pm$2.6 & 66.3$\pm$7.9 & ?92.7$\pm$4.0 & ?96.3$\pm$8.9 \\ \hline
\end{tabular}
\end{center}
\vglue-2mm
\caption{Updated \jpsi, \psip\ and Drell-Yan production
  cross-sections, in p-nucleus collisions at 450~GeV, as extracted
  from the ``HI~96/98'' and ``LI~98/00'' data sets, in the $-0.5<
  y_{\rm cm}<0.5$ and $-0.5<\cos \theta_{CS} <0.5$ phase space
  window.}
\label{oldpacs}
\end{table}

Table~\ref{tab:jpppsim} shows the results of each of the three data
sets, independently, in what concerns the nuclear dependences of the
\jpsi\ and \psip\ production cross-sections, as determined using the
detailed Glauber formalism.  It also shows, in the last line, the
results obtained from a simultaneous fit to the three data sets, with
a common absorption cross-section, $\sigma^{\rm G}_{\rm abs}$, and two
different normalisations, accounting for the different energies and
rapidity domains (we remind that the 400~GeV data is analysed in the
window $-0.425<y_{\rm cm}<0.575$).  The ``HI~2000'' absolute
cross-section systematic errors, affecting all targets in the same
way, are not considered in the global fit, since the 400~GeV data has
its own free normalisation.

\begin{table}[ht]
\begin{center}
\catcode`?=\active \def?{\kern\digitwidth}
\begin{tabular}{c|ccc|ccc}
\hline
\rule{0pt}{0.45cm}
& $\sigma_0^{\rm G}(\rm J/\psi)$  & $\sigma^{\rm G}_{\rm abs}(\rm J/\psi)$ & $\chi^2$/ndf
& $\sigma_0^{\rm G}(\psi^\prime)$ & $\sigma^{\rm G}_{\rm abs}(\psi^\prime)$ & $\chi^2$/ndf \\
& (nb/nuc) & (mb) & & (pb/nuc) & (mb) & \\
\hline
 HI 96/98    & 5.5 $\pm$ 0.2 & 4.4 $\pm$ 1.0 & 0.9 & 101 $\pm$ 5 & ?7.6 $\pm$ 1.4 & 1.3 \\ 
 LI 98/00    & 5.7 $\pm$ 0.3 & 4.1 $\pm$ 1.4 & 0.6 & 100 $\pm$ 7 & ?5.7 $\pm$ 2.0 & 0.6 \\
 HI 2000     & 5.1 $\pm$ 0.1 & 4.6 $\pm$ 0.6 & 0.4 & 101 $\pm$ 6 & 10.1 $\pm$ 1.6 & 1.7 \\ \hline
 450~GeV     & 5.6 $\pm$ 0.1 & \multirow{2}{*}{4.5 $\pm$ 0.5} & \multirow{2}{*}{0.7}
             & 106 $\pm$ 4   & \multirow{2}{*}{8.3 $\pm$ 0.9} & \multirow{2}{*}{1.3} \\
 400~GeV     & 5.1 $\pm$ 0.1 &             &     & ?95 $\pm$ 4 &      \\ \hline
\end{tabular}
\end{center}
\vglue-2mm
\caption{Nuclear dependence parameters of the \jpsi\ and \psip\ 
production cross-sections, from each of the three data sets and 
from a global fit to all of them.}
\label{tab:jpppsim}
\end{table}

\begin{figure}[ht]
\centering
\begin{tabular}{cc}
\resizebox{0.48\textwidth}{!}{%
\includegraphics*{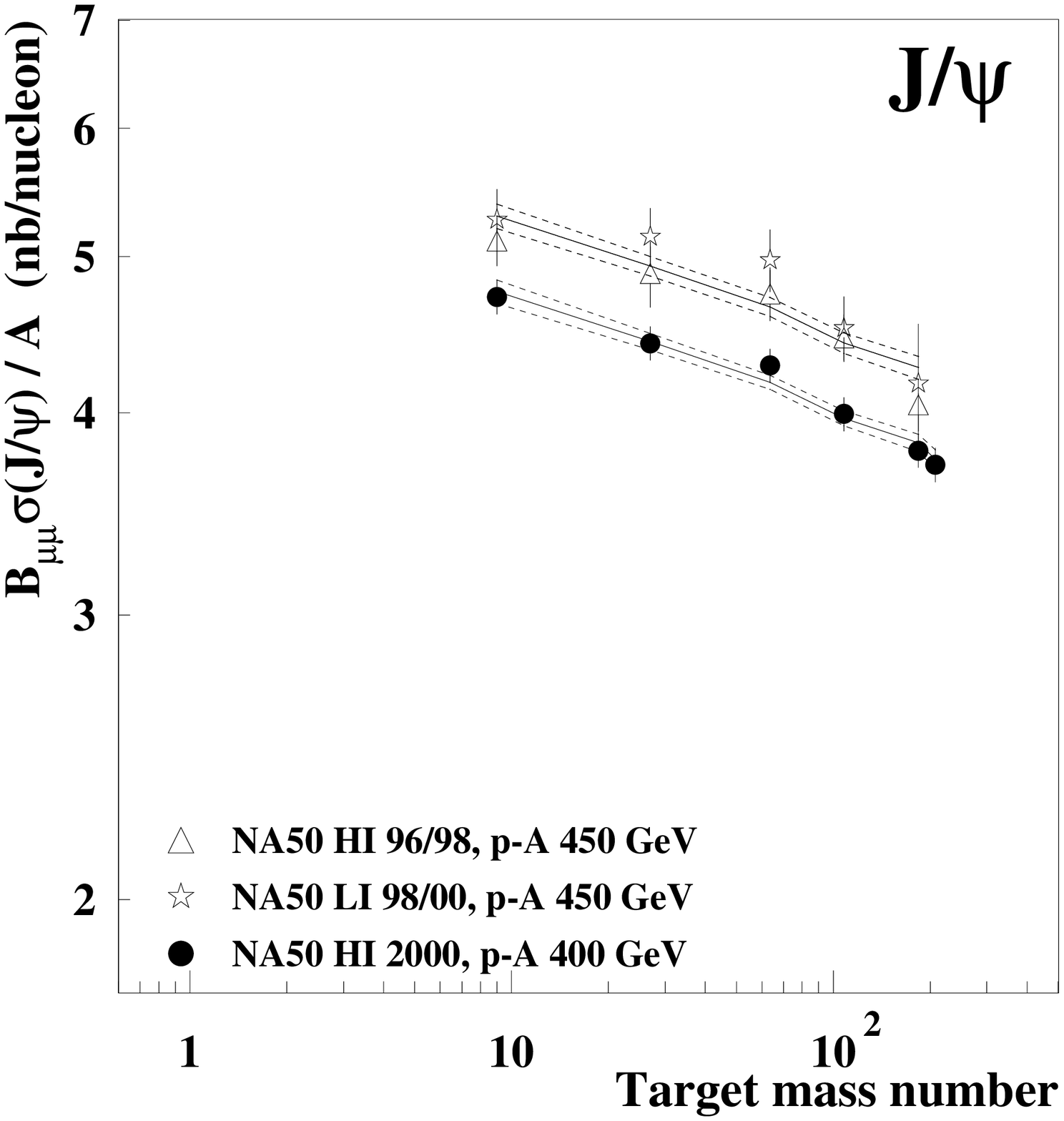}}
&
\resizebox{0.48\textwidth}{!}{%
\includegraphics*{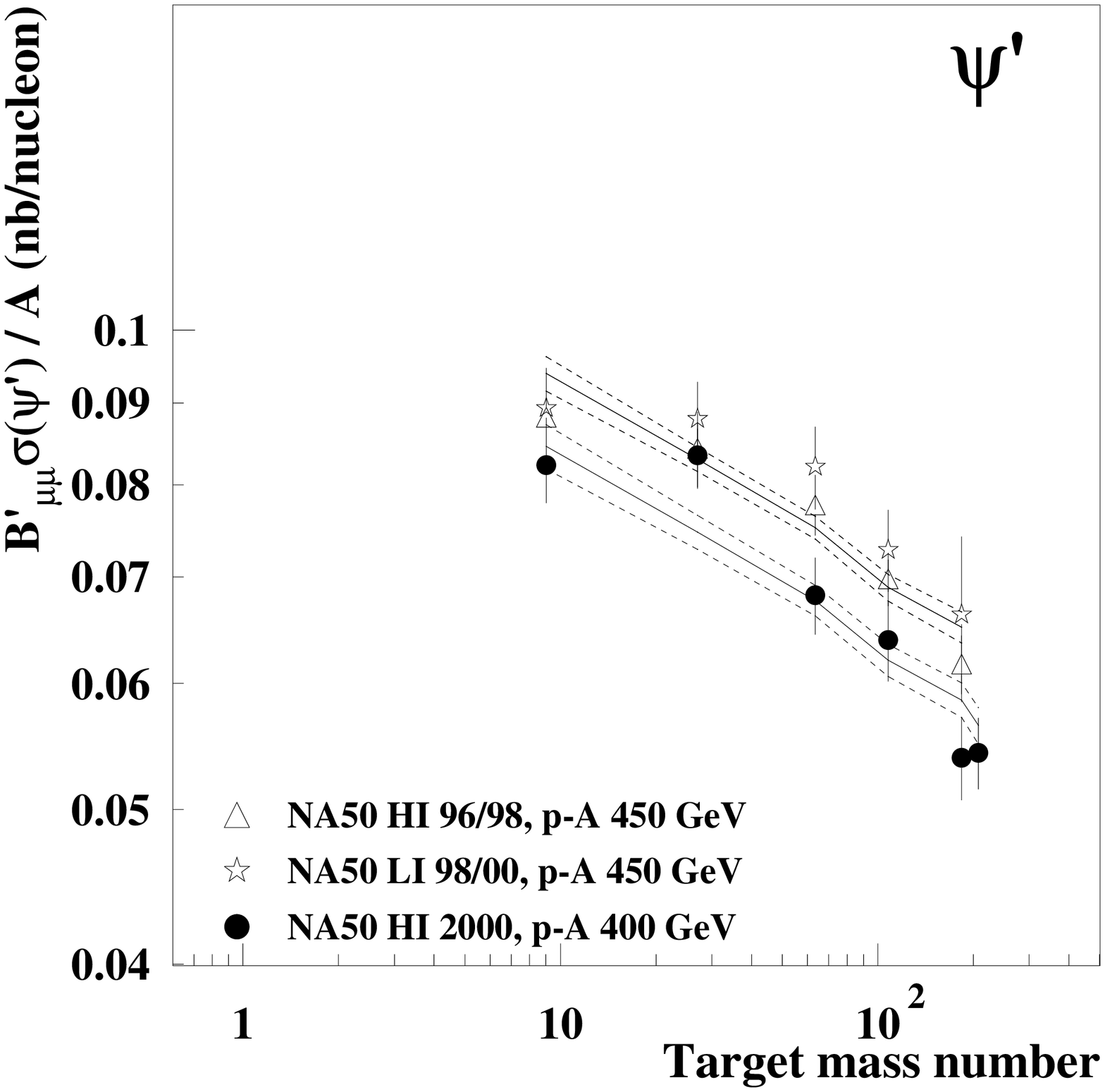}}
\end{tabular}
\caption{\jpsi\ (left) and \psip\ (right) cross-sections from
the three different NA50 data sets.  The full line represents the
result of a simultaneous Glauber fit with a common absorption
cross-section and two independent normalisations.  The error bands
reflect the absorption cross-section and normalisation uncertainties.}
\label{glbsimfit}
\end{figure}

Figure~\ref{glbsimfit} shows together all the \jpsi\ (left) and \psip\
(right) absolute production cross-sections measured by NA50 in
p-nucleus collisions, as collected in Tables~\ref{abs-xsec}
and~\ref{oldpacs}.  The curves are the result of the simultaneous
Glauber fits to all the data points and the error bands represent the
uncertainties in the absorption cross-section and in the
normalisations (see Table~\ref{tab:jpppsim} for the numerical values).

If instead of using the full Glauber calculation we use the approximate 
$\rho L$ parameterization, $\sigma_0$ does not change but the 
$\sigma^{\rho L}_{\rm abs}$ values are smaller, by $\sim$\,10\,\%
for the \jpsi\ and by $\sim$\,15\,\% for the \psip.  

The $\rho L$ parameterization is particularly useful to determine the 
relative absorption cross-section of the \psip\ with respect to that of the 
\jpsi, $\Delta \sigma^{\rho L}_{\rm abs}$. In the same way, the $\alpha$ 
parameterization is also used to quantify the relative charmonia absorption 
through the study of the $\Delta \alpha$ parameter.

\begin{table}[h!]
\begin{center}
\catcode`?=\active \def?{\kern\digitwidth}
\begin{tabular}{c|ccc|ccc}
\hline

& $\sigma_0^{\rho L}(\psi^\prime/{\rm J/\psi})$
& $\Delta \sigma^{\rho L}_{\rm abs}$ & $\chi^2$/ndf 
& $\sigma_0^{\alpha}(\psi^\prime/{\rm J/\psi})$
& $\Delta \alpha$ & $\chi^2$/ndf \\
& (\%) & (mb) & & (\%) & & \\
\hline
 HI 96/98 & 1.83$\pm$0.06 & 2.6$\pm$0.6 & 0.5 & 1.95$\pm$0.09 & -0.045$\pm$0.011 & 0.7 \\
 LI 98/00 & 1.79$\pm$0.08 & 1.9$\pm$1.0 & 0.4 & 1.87$\pm$0.13 & -0.032$\pm$0.017 & 0.5 \\
 HI 2000  & 1.98$\pm$0.10 & 4.2$\pm$1.0 & 1.7 & 2.21$\pm$0.17 & -0.076$\pm$0.018 & 1.9 \\ \hline
\multirow{2}{*}{Global} & 
\multirow{2}{*}{1.85$\pm$0.04} & \multirow{2}{*}{2.8$\pm$0.5}      & \multirow{2}{*}{0.9} &
\multirow{2}{*}{1.98$\pm$0.07} & \multirow{2}{*}{-0.048$\pm$0.008} & \multirow{2}{*}{1.0} \\
& & & & & & \\ \hline
\end{tabular}
\end{center}
\vglue-2mm
\caption{Relative nuclear absorption of the \psip\ with respect to the
  \jpsi, obtained from the ${\rm
  B^\prime_{\mu\mu}\sigma(\psi^\prime)}/{\rm B_{\mu\mu} \sigma(\psi)}$
  ratios, using the $\rho L$ and $\alpha$ parameterizations.}
\label{tab:ppoverjpsim}
\end{table}

Table~\ref{tab:ppoverjpsim} shows the $\Delta \sigma^{\rho L}_{\rm
abs}$ and $\Delta \alpha$ values independently obtained from each 
of the three data sets and also when we perform a simultaneous fit 
to all the data, assuming that the ratio between the \psip\ and the 
\jpsi\ production cross-sections does not change from 400 to 450~GeV.  
We can see that the values given by the global fit, 
$\Delta \sigma^{\rho L}_{\rm abs}=2.8\pm0.5$~mb and 
$\Delta \alpha=-0.048\pm0.008$, are essentially the same as the ones 
determined from the high statistics ``HI~96/98'' data set, on its own.

From the global fits leading to Table~\ref{tab:jpppsim}, we see that
the ratio between the charmonia production cross-sections at 400~GeV,
in the $-0.425<y_{\rm cm}<0.575$ window, and at 450~GeV, in the
$-0.5<y_{\rm cm}<0.5$ window, is 0.90 (with a 3\,\% relative error
bar).  This value is in perfect agreement with the expectation of a
NLO QCD calculation made in the Color Evaporation Model 
framework~\cite{nlorvogt}.  Using the ``Schuler 
parameterization''~\cite{schuler}, which provides a phenomenological 
description of \jpsi\ production cross-sections as a function of $\sqrt{s}$, 
we would expect $0.933\pm0.008$, a value which includes the 
corresponding changes in the \xf\ distributions and that is quite 
close to the result derived from our analysis.

As already briefly mentioned at the end of the previous section, it is
interesting to study the \jpsi\ and \psip\ nuclear dependences using
the ratios of production cross-sections $B_{\mu\mu}\sigma({\rm
J}/\psi)/\sigma(DY_{2.9-4.5})$ and
$B^\prime_{\mu\mu}\sigma(\psi^\prime)/\sigma(DY_{2.9-4.5})$.  In what
concerns the ``HI~2000'' data set, these values have already been
presented in Table~\ref{rat-xsec}.  The corresponding values for the
two other data sets are collected in Table~\ref{oldratios}.  They were
obtained from the analyses described in the
references~\cite{pcpaper,rspaper,RubenPhD} and corrected, where
appropriate, as explained above.

\begin{table}[ht]
\begin{center}
\catcode`?=\active \def?{\kern\digitwidth}
\begin{tabular}{c|cc|cc|cc}
\hline
\rule{0pt}{0.45cm}
& \multicolumn{2}{c|}{$\frac{B_{\mu\mu}\sigma({\rm J}/\psi)}{\sigma(DY_{2.9-4.5})}$}
& \multicolumn{2}{c|}{$\frac{B^\prime_{\mu\mu}\sigma(\psi^\prime)}{\sigma(DY_{2.9-4.5})}$}
& \multicolumn{2}{c}{$\frac{B^\prime_{\mu\mu}\sigma(\psi^\prime)}{B_{\mu\mu}\sigma({\rm
      J}/\psi)}$}
\\ \hline
    &   HI 96/98   &   LI 98/00   &     HI 96/98    &     LI 98/00    &   HI 96/98    &    LI 98/00   \\ \hline 
 Be & 54.7$\pm$1.5 & 54.6$\pm$1.8 & 0.946$\pm$0.039 & 0.932$\pm$0.051 & 1.73$\pm$0.05 & 1.70$\pm$0.07 \\  
 Al & 50.8$\pm$1.2 & 49.9$\pm$1.7 & 0.876$\pm$0.030 & 0.859$\pm$0.044 & 1.73$\pm$0.04 & 1.71$\pm$0.07 \\
 Cu & 46.7$\pm$1.0 & 51.0$\pm$2.0 & 0.768$\pm$0.024 & 0.848$\pm$0.044 & 1.64$\pm$0.03 & 1.66$\pm$0.05 \\
 Ag & 47.2$\pm$1.0 & 45.5$\pm$1.3 & 0.740$\pm$0.021 & 0.735$\pm$0.034 & 1.57$\pm$0.03 & 1.61$\pm$0.05 \\
 W  & 43.7$\pm$0.9 & 43.6$\pm$1.7 & 0.666$\pm$0.020 & 0.665$\pm$0.044 & 1.53$\pm$0.04 & 1.52$\pm$0.07 \\ \hline
\end{tabular}
\end{center}
\vglue-2mm
\caption{Ratios of \jpsi, \psip\ and Drell-Yan production
cross-sections, in p-A collisions at 450~GeV, as extracted from the
``HI~96/98'' and ``LI~98/00'' NA50 data sets, in the $-0.5<y_{\rm
cm}<0.5$ and $-0.5<\cos \theta_{CS} <0.5$ kinematical window.}
\label{oldratios}
\end{table}

From the cross-section ratios of Tables~\ref{rat-xsec}
and~\ref{oldratios} we can extract the nuclear absorption parameters,
using the Glauber formalism, as we have done from the absolute
cross-sections (in Table~\ref{tab:jpppsim}).  The results are
summarised in Table~\ref{tab:ratiossim}, where the last line shows the
values resulting from a global fit to the three data sets.
The derivation of the charmonia nuclear absorption parameters using
the ratios of cross-sections, $B_{\mu\mu}\sigma({\rm
J}/\psi)/\sigma(DY_{2.9-4.5})$ and
$B^\prime_{\mu\mu}\sigma(\psi^\prime)/\sigma(DY_{2.9-4.5})$, is very
important to establish the ``normal nuclear absorption curve'' in
heavy-ion collisions, where the suppression pattern of the \jpsi\ and
\psip\ resonances is determined, for each centrality bin, using the
ratio of their production yields to the corresponding yield of Drell-Yan 
events.

\begin{table}[ht]
\begin{center}
\catcode`?=\active \def?{\kern\digitwidth}
\begin{tabular}{c|ccc|ccc}
\hline
\rule{0pt}{0.45cm}
& $\sigma_0^{\rm G}(\rm J/\psi\,/\,{\rm DY})$  & $\sigma^{\rm G}_{\rm abs}(\rm J/\psi)$ & $\chi^2$/ndf
& $\sigma_0^{\rm G}(\psi^\prime\,/\,{\rm DY})$ & $\sigma^{\rm G}_{\rm abs}(\psi^\prime)$ & $\chi^2$/ndf \\
& (nb/nuc) & (mb) & & (nb/nuc) & (mb) & \\
\hline
 HI 96/98 & 57.7 $\pm$ 1.7 & 4.3 $\pm$ 0.7 & 1.5 & 1.07 $\pm$ 0.05 & 7.8 $\pm$ 1.2 & 0.8 \\
 LI 98/00 & 58.2 $\pm$ 2.3 & 4.4 $\pm$ 1.0 & 0.9 & 1.05 $\pm$ 0.07 & 6.8 $\pm$ 1.7 & 0.8 \\
 HI 2000  & 57.2 $\pm$ 3.4 & 3.4 $\pm$ 1.2 & 0.4 & 1.14 $\pm$ 0.11 & 8.8 $\pm$ 2.3 & 0.9 \\
\hline
 450~GeV  & 57.5 $\pm$ 1.3 & \multirow{2}{*}{4.2 $\pm$ 0.5} & \multirow{2}{*}{0.7} 
          & 1.07 $\pm$ 0.04 & \multirow{2}{*}{7.7 $\pm$ 0.9} & \multirow{2}{*}{0.7} \\
 400~GeV  & 59.3 $\pm$ 1.8 & & & 1.09 $\pm$ 0.05 & & \\ \hline
\end{tabular}
\end{center}
\vglue-2mm
\caption{Nuclear dependence parameters of the \jpsi\ and \psip\ 
resonances, using cross-section ratios, from each of the three data 
sets and from a global fit to all of them.}
\label{tab:ratiossim}
\end{table}

It is important to note that the \jpsi\ absorption cross-section
derived from the \jpsi\,/\,DY ratios, $\sigma^{\rm G}_{\rm abs}(\rm
J/\psi) = 4.2\pm0.5$~mb, is in excellent agreement with the value
derived from the absolute production cross-sections, $4.5\pm0.5$~mb
(Table~\ref{tab:jpppsim}). Also the \psip\ values obtained in the two
alternative approaches are in very good agreement, $\sigma^{\rm
G}_{\rm abs}(\psi^\prime) = 7.7\pm0.9$~mb (Table~\ref{tab:ratiossim})
and $8.3\pm0.9$~mb (Table~\ref{tab:jpppsim}). 
When defining the normal nuclear absorption curve, as a
function of centrality, in heavy-ion collisions, to compare with the
\jpsi\ suppression pattern obtained in those collisions through the
\jpsi\,/\,DY cross-section ratio, the value $4.2\pm0.5$~mb should be
used, for consistency.
This $\sigma^{\rm G}_{\rm abs}$ value is in agreement with previous
published results \cite{pcpaper,PbPb2000} but it should be noticed 
that when NA51 pp and pD charmonia measurements~\cite{NA51} were 
included in the Glauber fit \cite{PbPb2000}, a 10\% systematic 
error had to be added due to the different analysis techniques 
used in NA51 data. The inclusion of the pp and pD values in the Glauber 
fit has a strong influence on the derived value of $\sigma^{\rm G}_{\rm abs}$ 
and since a coherent analysis including NA51 points cannot be performed
any more due to raw data unavailability, these results were excluded from 
the analysis reported here.

As a final remark, the $\sigma^{\rm G}_{\rm abs}$ energy and kinematical 
dependence was also investigated using other p-A production cross-sections, 
published by experiments with similar spectrometers: NA3~\cite{Bad83} and 
NA38~\cite{MAbreu}, at $E_{\rm lab}=200$~GeV. These values, fitted with the 
full Glauber model, lead to $\sigma^{\rm G}_{\rm abs}(\rm J/\psi) = 3.3\pm3.0$~mb,
which, within the present precision, is compatible with our present measurement.
The inclusion of these points in the global Glauber fit 
using \jpsi\ cross-sections does not change the final $\sigma^{\rm G}_{\rm abs}$ 
value. Nevertheless, the energy dependence of the absorption cross-section remains, 
today, an open question.

\section{Summary and conclusions}

This paper presents the \jpsi, \psip\ and Drell-Yan production
cross-sections in proton-nucleus collisions, as extracted from data
collected by NA50 in year 2000, with 400~GeV protons and six different
targets: Be, Al, Cu, Ag, W and Pb.  The frequent rotation of the
target material facing the beam significantly decreases the
time-dependent systematic uncertainties related to beam stability,
trigger system, etc.  The accumulated statistics, however, suffers
from the very short beam time available, which affects in particular
the \psip\ and Drell-Yan production results.

The nuclear dependence of the charmonium states production is studied, 
as a function of the target nucleus, with three different descriptions: 
the $\alpha$ parameterization, the detailed Glauber formalism and its
``$\rho L$'' approximation. The \jpsi\ and \psip\ absorption cross-sections 
obtained with the Glauber model, from the 400\,GeV data sets, are: 
\mbox{$\sigma^{\rm G}_{\rm abs}(\rm J/\psi)=4.6\pm0.6$~mb} and 
\mbox{$\sigma^{\rm G}_{\rm abs}(\psi^\prime)=10.0\pm1.5$~mb}.

A detailed comparison with previous NA50 p-A data samples, collected
at 450~GeV with five different nuclear targets, shows that all data
can be described with common absorption cross-sections. 
The ``HI~2000'' data set is particularly important in the determination 
of $\sigma_{\rm abs}(\rm J/\psi)$ from the study of the absolute production 
cross-sections, given the lower systematic uncertainties on the beam 
normalisations. On the other hand, the two data sets previously collected by 
NA50 have more statistics and, therefore, lead to more accurate measurements 
of the $B_{\mu\mu}\sigma({\rm J}/\psi)/\sigma(DY_{2.9-4.5})$ and
$B^\prime_{\mu\mu}\sigma(\psi^\prime)/\sigma(DY_{2.9-4.5})$
cross-section \emph{ratios}, which are insensitive to the beam normalisation.  
Global fits to the cross-section values give: $\sigma^{\rm G}_{\rm
abs}(\rm J/\psi) = 4.5\pm0.5$~mb and $\sigma^{\rm G}_{\rm
abs}(\psi^\prime)=8.3\pm0.9$~mb.  The corresponding results from the
$B_{\mu\mu}\sigma({\rm J}/\psi)/\sigma(DY_{2.9-4.5})$ and
$B^\prime_{\mu\mu}\sigma(\psi^\prime)/\sigma(DY_{2.9-4.5})$
cross-section ratios are in excellent agreement: $\sigma^{\rm G}_{\rm
abs}(\rm J/\psi) = 4.2\pm0.5$~mb and $\sigma^{\rm G}_{\rm
abs}(\psi^\prime)=7.7\pm0.9$~mb. The fact that the two alternative ways of 
deriving the charmonia absorption cross-sections, through absolute production 
cross-sections or through their ratios, are in such good agreement (furthermore, 
being mostly determined from different data sets), is an independent and accurate 
indication that, at our energies and in our kinematical phase space window, 
Drell-Yan production is not sensitive to nuclear effects, including initial state 
effects such as nuclear modifications of the quark distribution functions and 
energy loss of the incident nucleons. It would be interesting to see if a similar 
study made as a function of the dimuon \pt\ or \xf\ would modify this observation.

For consistency reasons, the values 
$\sigma^{\rm G}_{\rm abs}(\rm J/\psi) = 4.2\pm0.5$~mb 
and $\sigma^{\rm G}_{\rm abs}(\psi^\prime)=7.7\pm0.9$~mb should be 
used in the comparison with the cross-section ratios
measured in heavy ion collisions. A significantly stronger absorption of 
the \psip\ resonance with respect to the \jpsi\ is seen through the study 
of the relative production yields between the two charmonium states, in our 
phase space window.

The results presented in this paper were obtained assuming that
final state nuclear absorption of the charmonium states, as
calculated through detailed Glauber calculations with a single free
parameter, $\sigma^{\rm G}_{\rm abs}$, is the only nuclear effect
influencing the production of these resonances in p-nucleus
collisions.  If future measurements (for instance, of open charm
production cross-sections in p-nucleus collisions) demonstrate
the existence of nuclear effects (such as anti-shadowing) on the gluon
distribution functions at SPS energies, the numerical values given
in this paper should be seen as effective values, resulting from the
convolution of initial and final state nuclear effects.


\section*{Acknowledgements}

This work was partially supported by the Funda\c{c}\~ao para a
Ci\^encia e a Tecnologia, Portugal.

\end{document}